\documentclass[journal,twoside,web]{IEEEtran}
\usepackage{amsmath,amsfonts}
\usepackage{algorithm}
\usepackage{array}
\usepackage[caption=false,font=normalsize,labelfont=sf,textfont=sf]{subfig}
\usepackage{textcomp}
\usepackage{stfloats}
\usepackage{url}
\usepackage{verbatim}
\usepackage{graphicx}
\usepackage{cite}

\usepackage{booktabs}
\usepackage{capt-of}

\usepackage{listings}		
\usepackage{enumerate}	
\usepackage{hyperref}
\newcommand{\envleftshift}{}
\renewcommand{\envleftshift}{0in} 	% Use this for usual paragraph indentation
\newtheorem{definition}{\hspace*{-\envleftshift}Definition}[section]

\newtheorem{lemma}{\hspace*{-\envleftshift}Lemma}[section]
\newtheorem{proposition}{\hspace*{-\envleftshift}Proposition}[section]
\newtheorem{theorem}{\hspace*{-\envleftshift}Theorem}[section]

\newtheorem{remark}{\hspace*{-\envleftshift}Remark}[section]

\usepackage{color,soul}

\usepackage[table]{xcolor}

\usepackage{gensymb}
\usepackage{amssymb}

\usepackage{mathtools}
\usepackage{mathrsfs}

\usepackage{tensor}

\usepackage{float}
\usepackage{placeins} 

\usepackage{hhline} 		% Custom horizontal lines in tables
\usepackage{multirow}		% Multirow feature in tables
\usepackage{chngpage}
\usepackage{lscape} 		% Turn page contents to landscape
\usepackage{tablefootnote} 	% For footnotes in tables. Use \tablefootnote{} command

\usepackage{tikz}
\usetikzlibrary{cd}				% For commuting diagrams

\usepackage{algpseudocode}

\usepackage{mdframed}

\newcommand\Tstrut{\rule{0pt}{2.6ex}}       % "top" strut
\newcommand{\relpath}{}
\newcommand{\relpathone}{}
\newcommand{\relpathtwo}{}

\newdimen\figsize
\newcommand{\lightshade}{20}
\usepackage{accents}

\DeclareMathOperator{\vecc}{\textnormal{vec}}
\DeclareMathOperator{\mat}{\textnormal{mat}}
\DeclareMathOperator{\svec}{\textnormal{svec}}
\DeclareMathOperator{\smat}{\textnormal{smat}}
\DeclareMathOperator{\skron}{\underline{\otimes}}
\DeclareMathOperator{\skrons}{\underline{\oplus}}

\DeclareMathOperator{\GL}{\textnormal{GL}}
\DeclareMathOperator{\rank}{\textnormal{rank}}

\DeclareMathOperator{\diag}{\textnormal{\texttt{diag}}}

\usepackage{bibentry}
\nobibliography*

\usepackage{tikz}				% For block diagrams

\usetikzlibrary{arrows,decorations,backgrounds,arrows.meta}
\usetikzlibrary{calc}
\usetikzlibrary{bending}
\usetikzlibrary{angles,positioning,intersections,quotes,decorations.markings}
\usetikzlibrary{matrix}

\usetikzlibrary{decorations.pathreplacing,decorations.markings,hobby}
\tikzset{
  on each segment/.style={
    decorate,
    decoration={
      show path construction,
      moveto code={},
      lineto code={
        \path [#1]
        (\tikzinputsegmentfirst) -- (\tikzinputsegmentlast);
      },
      curveto code={
        \path [#1] (\tikzinputsegmentfirst)
        .. controls
        (\tikzinputsegmentsupporta) and (\tikzinputsegmentsupportb)
        ..
        (\tikzinputsegmentlast);
      },
      closepath code={
        \path [#1]
        (\tikzinputsegmentfirst) -- (\tikzinputsegmentlast);
      },
    },
  },
  mid arrow/.style={postaction={decorate,decoration={
        markings,
        mark=at position .5 with {\arrow[#1]{stealth}}
      }}},
}
\tikzset{
  multi arrow/.style n args={3}{
  		postaction={decorate,decoration={
        markings,
        mark = between positions #1 and #2 step #3 with {\arrow{stealth}}
      }}},      
      multi arrow/.default={0.1}{0.9}{0.4}		% Default args
}
\tikzset{
  multi arrow reversed/.style n args={3}{
  		postaction={decorate,decoration={
        markings,
        mark = between positions #1 and #2 step #3 with {\arrowreversed{stealth}}
      }}},      
      multi arrow reversed/.default={0.1}{0.9}{0.4}		% Default args
}
\tikzset{
  nyquist forward/.style={
  		line width = \Linewidthbf,
  		postaction = {multi arrow}
  		}     
}
\tikzset{
  nyquist reverse/.style={
  		line width = \Linewidthbf,
  		dashed,
  		postaction = {multi arrow reversed},
  		}     
}
\tikzset{
	pics/rangle/.style n args={4}{
  	code = {	
		\pgftransformrotate{#1}		% Rotate about +(0,0) by argument 1
		\pgftransformscale{#2}		% Scale wrt +(0,0) by argument 2
		\coordinate (p1) at (0,0);
		\coordinate (p2) at (-1,0);
		\coordinate (p3) at (-1,1);
		\coordinate (rp1) at ([shift={(#4,0.0)}]p2);
		\coordinate (rp3) at ([shift={(0.0,#4)}]p2);
		\coordinate (rp2) at (rp1 |- rp3);
		\pgftransformreset			% Reset pgf transformations
		\draw [#3] (p1)				% Apply style command of argument 3						
		coordinate (-start)			% Coord: start
		-- (p2) 
		coordinate (-corner)		% Coord: corner	
		-- (p3)	
		coordinate (-end)			% Coord: end
		; 			% END
		\draw [solid,thick]							
		(rp1)
		-- (rp2) 
		-- (rp3)
		; 			% END
  	}		
  	},
  	pics/rangle/.default={0}{1}{}{0.3}		% Default args     
}

\makeatletter
\newcommand\currentcoordinate{\the\tikz@lastxsaved,\the\tikz@lastysaved}
\makeatother

\newdimen\Linewidth \Linewidth = 0.2mm
\newdimen\Linewidthbf \Linewidthbf = 0.6mm
\newdimen\Linewidthtick \Linewidthtick = 0.3mm
\newdimen\Linewidthpz \Linewidthpz = 0.4mm
\newdimen\Shiftarrowbf \Shiftarrowbf = 1.5mm
\newdimen\Shiftlabely \Shiftlabely = -5mm
\newdimen\Shiftlabelx \Shiftlabelx = 8mm
\newcommand{\Arrowbf}{{latex[width=2mm]}}

\tikzstyle{zero}=[circle,draw=black,fill=white,line width = \Linewidthpz,inner sep=0pt, minimum size=1.4ex]
\tikzstyle{point}=[circle,draw=black,fill=black,line width = \Linewidthpz,inner sep=0pt, minimum size=0.5ex]

\pgfdeclarelayer{background}
\pgfdeclarelayer{foreground}
\pgfsetlayers{background,main,foreground}

\newcommand\copyrighttext{%
	\footnotesize \textcopyright 2023 IEEE. Personal use of this material is permitted. Permission from IEEE must be obtained for all other uses, in any current or future media, including reprinting/republishing this material for advertising or promotional purposes, creating new collective works, for resale or redistribution to servers or lists, or reuse of any copyrighted component of this work in other works.
  }
\newcommand\copyrightnotice{%
\begin{tikzpicture}[remember picture,overlay]
\node[anchor=south,yshift=10pt] at (current page.south) {\fbox{\parbox{\dimexpr\textwidth-\fboxsep-\fboxrule\relax}{\copyrighttext}}};
\end{tikzpicture}%
}

\newcommand{\TNNLSdEIRLCitation}{BA_Wallace_J_Si_dEIRL_TNNLS_arXiv:2023}

\newcommand{\TNNLSdEIRLSecESSetup}{VII-A}

\begin{document}

% *************************************************************************
%
% TITLE, FRONTMATTER
%
% *************************************************************************

%\iffalse			% Comment this line and \fi to include section

% ************************************************************************
% ************************************************************************
% ************************************************************************
%
% TITLE, FRONTMATTER
%
% ************************************************************************
% ************************************************************************
% ************************************************************************

% ***********************
%
% TITLE AND THANKS
%

\title{
Modulation-Enhanced Excitation for Continuous-Time Reinforcement Learning via Symmetric Kronecker Products
}

\author{
    Brent A. Wallace and Jennie Si,~\IEEEmembership{Fellow,~IEEE}
    \thanks{This work was supported in part by the NSF under Grants 1808752 and 2211740. Brent A. Wallace was also supported by the NSF under Graduate Research Fellowship Grant 026257-001.}
    \thanks{Brent A. Wallace (corresponding author) and Jennie Si are with the Department of Electrical, Computer \& Energy Engineering, Arizona State University, Tempe, AZ 85287 USA (e-mail: bawalla2@asu.edu; si@asu.edu).}
}   % END \AUTHOR BLOCK

% ***********************
%
% PAPER HEADER
%

%\markboth{IEEE Transactions on Automatic Control,~Vol.~ZZ, No.~ZZ, January~2023}%

% ***********************
%
% IEEE PUBLICATION ID
%

%\IEEEpubid{0000--0000/00\$00.00~\copyright~2022 IEEE}

\maketitle

% ************************************************************************
%
% ARXIV -- COPYRIGHT NOTICE
%
% ************************************************************************

\copyrightnotice

%\FloatBarrier

%\fi				% Comment this line and \iffalse to include section

% *************************************************************************
%
% ABSTRACT
%
% *************************************************************************

%\iffalse			% Comment this line and \fi to include section

% ************************************************************************
% ************************************************************************
% ************************************************************************
%
% ABSTRACT
%
% ************************************************************************
% ************************************************************************
% ************************************************************************

\begin{abstract}

This work introduces new results in continuous-time reinforcement learning (CT-RL) control of affine nonlinear systems to address a major algorithmic challenge due to a lack of persistence of excitation (PE).
This PE design limitation has previously stifled CT-RL numerical performance and prevented these algorithms from achieving control synthesis goals.
Our new theoretical developments in symmetric Kronecker products enable a proposed modulation-enhanced excitation (MEE) framework to make PE significantly more systematic and intuitive to achieve for real-world designers. 
MEE is applied to the suite of recently-developed excitable integral reinforcement learning (EIRL) algorithms, yielding a class of enhanced high-performance CT-RL control design methods which, due to the symmetric Kronecker product algebra, retain EIRL's convergence and closed-loop stability guarantees.
Through numerical evaluation studies, we demonstrate how our new MEE framework achieves substantial improvements in conditioning when approximately solving the Hamilton-Jacobi-Bellman equation to obtain optimal controls. We use an intuitive example to provide insights on the central excitation issue under discussion, and we demonstrate the effectiveness of the proposed procedure on a real-world hypersonic vehicle (HSV) application.

\end{abstract}


%\FloatBarrier

%\fi				% Comment this line and \iffalse to include section

% *************************************************************************
%
% INTRODUCTION
%
% *************************************************************************

%\iffalse			% Comment this line and \fi to include section

% ************************************************************************
% ************************************************************************
% ************************************************************************
%
% INTRODUCTION
%
% ************************************************************************
% ************************************************************************
% ************************************************************************

\section{Introduction \& Motivation}\label{sec:introduction}

% ************************************************************************
%
% BACKGROUND: DP, RL, ADP
%
% ************************************************************************

\IEEEPARstart{A}{daptive} dynamic programming (ADP) \cite{Werbos_RL_contr:book:1991, Werbos_ADP:incollection:1992, Si_Barto_Powell_Wunsch_approx_DP:book, Bertsekas_opt_contr:book} has proven a vital application of reinforcement learning (RL) \cite{Barto_Sutton_Anderson_RL:1983,Sutton_Barto_RL_opt_contr:book} to complex decision and control problems. ADP uses approximation and learning to solve the optimal control problem for both continuous-time (CT) and discrete-time (DT) dynamical systems, tackling the central ``curse of dimensionality" which has plagued the field of dynamic programming (DP) \cite{Bellman_DP:book} and limited applications in optimal control \cite{FL_Lewis_D_Vrabie_Syrmos_opt_contr:book}.

On one hand, review of DT-RL algorithms  \cite{Lewis_Vrabie_Vamvoudakis_RL_Contr_Overview:2012,Kiumarsi_KG_Vamvoudakis_H_Modares_FL_Lewis_RL_survey:2018} shows that they have demonstrated excellent stability, convergence, and approximation guarantees. For representative results, see, e.g.,  \cite{Mu_Wang_He_NDP:2017,Liu_Wei_PI_DT:2015,Guo_J_si_Liu_Mei_PI_approx:2018,Gao_Si_Wen_Li_Huang_prosthetics:2021,Al-Tamimi_FL_Lewis_Abu-Khalaf_VI_proof:2008,Liu_Sun_J_Si_Guo_Mei_dHDP_boundedness_result:2012}. 
DT-RL algorithms have also demonstrated great successes in addressing complex real-world control applications, such as robot position control \cite{Mondal_AA_Rodriguez_Manne_Das_BA_Wallace_wheeled_robot:2019,Mondal_BA_Wallace_AA_Rodriguez_wheeled_robot:2020}, power system stability enhancement \cite{Lu_J_Si_Xie_dHDP_power_sys:2008,Guo_Liu_J_Si_He_Harley_Mei_power_sys_stability:2015,Guo_Liu_J_Si_He_Harley_Mei_power_sys_freq_contr:2016}, helicopter stabilization, tracking, and reconfiguration control \cite{Enns_J_Si_Apache_NDP:2002,Enns_Si_helicopter:2003,Enns_J_Si_helicopter_rotor_failure:2003}, waste water treatment \cite{Yang_Cao_Meng_J_Si_wastewater:2021}, and wearable prostheses \cite{Wen_Liu_J_Si_Huang_ADP_transfemoral:2016,Wen_J_Si_Gao_Huang_Huang_prosthesis_lower_limb_ADP:2017,Wen_J_Si_Brandt_Gao_Wang_prosthesis_RL_contr:2019,Wu_Li_Yao_Liu_J_Si_Huang_prosthesis_RL_impedance_contr:2022,Li_Wen_Gao_J_Si_Huang_prosthesis_personalized_impedance:2022,Wu_Zhong_BA_Wallace_Gao_Huang_J_Si_prosthesis_symmetry:2022}.

On the other hand, CT-RL algorithms \cite{Vrabie_Lewis_IRL:2009,Vamvoudakis_Lewis_SPI:2010,Jiang_ZP_Jiang_Robust_ADP:2014,Bian_ZP_Jiang_ADP_VI:2021} have seen fewer theoretical developments and almost no applications successes when compared to their DT-RL counterparts. Recent comprehensive numerical analysis of prevailing ADP-based CT-RL algorithms \cite{BA_Wallace_J_Si_CT_RL_review:2022} shows that not only do they suffer from significant algorithm complexity issues, they also struggle with persistence of excitation (PE) as a central design limitation.
This fundamental limitation results in crippling numerical performance issues; in particular, poor conditioning of the underlying learning regression.
Altogether, these design limitations stifle the real-world synthesis performance of current CT-RL algorithms. We thus are still in search of formal CT-RL control design methods \cite{BA_Wallace_J_Si_CT_RL_review:2022}.
%and make them highly difficult for designers to use systematically \cite{BA_Wallace_J_Si_CT_RL_review:2022}.  
%
In response to this great PE issue, the authors in \cite{\TNNLSdEIRLCitation} develop a suite of excitable integral reinforcement learning (EIRL) algorithms, especially the decentralized variant dEIRL. The original dEIRL study rigorously proves convergence and closed-loop stability, and it demonstrates real-world synthesis guarantees \cite{\TNNLSdEIRLCitation}.

Although dEIRL has demonstrated significant reductions in conditioning relative to prior CT-RL methods, there is still a further underlying barrier to achieving PE \cite{\TNNLSdEIRLCitation}. In particular, previous empirical studies reveal that learning regression conditioning suffers due to physical constraints such as actuator saturations, high-frequency model uncertainties, and unit intermingling (e.g., m, m/s in translational loops and deg, deg/s in rotational loops).
These constraints force a gap between the excitation level permissible by the underlying physical process and the excitation level required for good algorithm numerics \cite{\TNNLSdEIRLCitation}.
Filling this gap requires new theoretical developments that can potentially elevate control synthesis relying on PE as a conceptual idea to a practically-useful tool for designers.

This work develops new properties of the symmetric Kronecker product, which compared to the standard Kronecker product \cite{Horn_matrix_analysis:book:1991,Brewer_Kronecker_products:1978} has received very little theoretical attention and has only been studied by a handful of important works \cite{Alizadeh_Haeberly_Overton_skron:1998,Todd_Toh_Tutuncu_skron:1998,Tuncel_Wolkowicz_skron:2005,Kalantarova_Tuncel_skron:2021,Kalantarova_PhD_thesis_skron:2019} (cf. Section \ref{sec:skron} for a summary of prior results/developments). 
These new algebraic results are essential to the proposed work; crucially, they ensure that MEE preserves dEIRL's convergence and closed-loop stability guarantees \cite{\TNNLSdEIRLCitation}. 
%Thus, MEE enables substantial conditioning improvements while preserving dEIRL's key performance guarantees. 
%
Furthermore, the symmetric Kronecker product results uncover substantial parallels in algebraic structure between dEIRL and the algebraic Lyapunov equation (ALE) approach of Kleinman's classical control framework \cite{Kleinman_AREs:1968}.

With these new theoretical developments, MEE allows designers to
apply first-principles insights of the dynamics to modulate the learning regression via nonsingular transformations of the state variables. When applied to the dEIRL algorithm, MEE may be used systematically in conjunction with dEIRL’s multi-injection and decentralization, comprising an unparalleled three-prong approach to tackle the CT-RL curses of dimensionality and conditioning \cite{BA_Wallace_J_Si_CT_RL_review:2022,\TNNLSdEIRLCitation}.

% *************************************************************************
% *************************************************************************
%
% SECTION: CONTRIBUTIONS
%
% *************************************************************************
% *************************************************************************

%\section{Contributions}

The contributions of this work are threefold: 
1) We develop a new modulation-enhanced excitation (MEE) framework to substantively address long-standing PE issues in CT-RL control. 
2) We apply MEE to the suite of EIRL algorithms, and we numerically demonstrate on a motivating example and real-world hypersonic vehicle study how MEE may be used as an intuitive design tool to yield significant numerical improvements while preserving EIRL's convergence and stability guarantees. 
3) To develop the MEE framework, we develop a new rectangular-matrix version of the symmetric Kronecker product and introduce the symmetric Kronecker sum operation, proving new fundamental algebraic and spectral results for both maps.

% *************************************************************************
%
% ORGANIZATION
%
% *************************************************************************

%\noindent\textbf{Organization.} 
%
The remainder of this work is organized as follows.
We first establish background and a formulation of the dEIRL algorithm in Section \ref{sec:algs_training}. We then motivate the need for the developed MEE framework via an intuitive example in Section \ref{sec:motivating_ex}. Subsequently, we derive the required symmetric Kronecker product algebra in Section \ref{sec:skron}, using this algebra to apply MEE to the dEIRL algorithm in Section \ref{sec:prescaling}. We demonstrate MEE in our evaluation studies of Section \ref{sec:ES}. Finally, we conclude this work with a discussion in Section \ref{sec:conclusion}.

%\FloatBarrier

%\fi				% Comment this line and \iffalse to include section

% *************************************************************************
%
% PROBLEM FORMULATION
%
% *************************************************************************

%\iffalse			% Comment this line and \fi to include section

% ************************************************************************
% ************************************************************************
% ************************************************************************
%
% PROBLEM FORMULATION
%
% ************************************************************************
% ************************************************************************
% ************************************************************************

\section{Background}\label{sec:algs_training}

% ************************************************************************
%
% NOTATION
%
% ************************************************************************

\noindent\textbf{Notation.}
We denote $\left< \cdot, \cdot \right>_{F}$ as the Frobenius inner product on $\mathbb{R}^{m \times n}$. Let $\otimes$, $\vecc$ denote the usual Kronecker product and vectorization operations, respectively, and $\mat = \vecc^{-1}$ \cite{Brewer_Kronecker_products:1978}. For any concepts pertaining to differential geometry, this work follows the notational conventions of the standard text \cite{Lee_diff_geom:book:2013}. For $n \in \mathbb{N}$, let $\GL(n) \subset \mathbb{R}^{n \times n}$ denote the (real) general linear group of square invertible $n \times n$ matrices. Let $\mathbb{S}^{n} \subset \mathbb{R}^{n \times n}$ denote the subspace of symmetric matrices, and let $\underline{n} = \tensor*[_{n}]{P}{_{2}} = \frac{n(n+1)}{2}$ denote the dimension of $\mathbb{S}^{n}$. 

% ************************************************************************
% ************************************************************************
% ************************************************************************
%
% SUBSECTION: PROBLEM FORMULATION
%
% ************************************************************************
% ************************************************************************
% ************************************************************************

\subsection{Problem Formulation}\label{sec:problem_formulation}

% *************************************************************************
%
% SYSTEM
%
% *************************************************************************

\noindent\textbf{System.} %\noindent\textbf{System.} 
We consider the continuous-time nonlinear time-invariant affine systems $(f, g)$ affording a decentralized dynamical structure with $N \in \mathbb{N}$ loops, which we present in the $N = 2$ case here for simplicity:
\begin{align}
	\left[
	\begin{array}{c}
		\dot{x}_{1}
		\\
		\dot{x}_{2}
	\end{array}
	\right]
	=
	\left[
	\begin{array}{c}
		f_{1}(x)
		\\
		f_{2}(x)
	\end{array}
	\right]	
	+
	\left[
	\begin{array}{cc}
		g_{11}(x) & g_{12}(x)
		\\
		g_{21}(x) & g_{22}(x)
	\end{array}
	\right]	
	\left[
	\begin{array}{c}
		u_{1}
		\\
		u_{2}
	\end{array}
	\right],		
	\label{eq:sys_nonlin_2x2}
\end{align}
\noindent where $x \in \mathbb{R}^{n}$ is the state vector, $u \in \mathbb{R}^{m}$ is the control vector, $x_{j} \in \mathbb{R}^{n_{j}}$, $u_{j} \in \mathbb{R}^{m_{j}}$ $(j = 1, 2 \triangleq N)$ with $n_{1} + n_{2} = n$, $m_{1} + m_{2} = m$. We assume that $f(0) = 0$, and that $f$ and $g$ are Lipschitz on a compact set $\Omega \subset \mathbb{R}^n$ containing the origin $x = 0$ in its interior. Define $g_{j} : \mathbb{R}^{n} \rightarrow \mathbb{R}^{n_{j} \times m}$,  $g_{j}(x) = \left[ \begin{array}{cc} g_{j1}(x) & g_{j2}(x) \end{array} \right]$.

% *************************************************************************
%
% LQR PROBLEM
%
% *************************************************************************

\noindent\textbf{LQR Problem.}
In the LQR problem, we consider the continuous-time linear time-invariant system $(A, B)$, partitioned analogously to the nonlinear system $(f, g)$ (\ref{eq:sys_nonlin_2x2}):
\begin{align}
	\left[
	\begin{array}{c}
		\dot{x}_{1}
		\\
		\dot{x}_{2}
	\end{array}
	\right]
	=
	\left[
	\begin{array}{cc}
		A_{11} & A_{12}
		\\
		A_{21} & A_{22}
	\end{array}
	\right]	
	\left[
	\begin{array}{c}
		x_{1}
		\\
		x_{2}
	\end{array}
	\right]
	+
	\left[
	\begin{array}{cc}
		B_{11} & B_{12}
		\\
		B_{21} & B_{22}
	\end{array}
	\right]	
	\left[
	\begin{array}{c}
		u_{1}
		\\
		u_{2}
	\end{array}
	\right].		
	\label{eq:sys_lin_2x2}
\end{align}
\noindent where $x \in \mathbb{R}^{n}$, $u \in \mathbb{R}^{m}$ are the state and control vectors, respectively, $A \in \mathbb{R}^{n \times n}$, and $B \in \mathbb{R}^{n \times m}$. We assume that $(A, B)$ is stabilizable \cite{AAR_multivariable:book}, and we denote $(A, B)$ as the linearization of $(f, g)$ (\ref{eq:sys_nonlin_2x2}). 
LQR considers the quadratic cost functional
\begin{align}
	J(x_{0})
	=
	\int_{0}^{\infty} (x^{T} Q x + u^{T} R u) \, d \tau,
	\label{eq:LQR_cost_funct}
\end{align}  
\noindent where $Q \in \mathbb{S}^{n}$, $Q \geq 0$ and $R \in \mathbb{S}^{m}$, $R > 0$ are the state and control penalty matrices, respectively. It is assumed that $(Q^{1/2}, A)$ is detectable \cite{AAR_multivariable:book}. For decentralization, we impose the block-diagonal cost structure 
\begin{align}
	Q 
	=
	\left[
	\begin{array}{cc}
		Q_{1} & 0
		\\
		0 & Q_{2}
	\end{array}
	\right]	,
	\quad
	R 
	=
	\left[
	\begin{array}{cc}
		R_{1} & 0
		\\
		0 & R_{2}
	\end{array}
	\right],	
    \label{eq:LQR_penalties_2x2}
\end{align}
where $Q_{j} \in \mathbb{S}^{n_{j}}$, $Q_{j} \geq 0$, and $R_{j} \in \mathbb{S}^{m_{j}}$, $R_{j} > 0$ $(j = 1, 2)$.
%
% *************************************************************************
%
% ALG OUTPUTS
%
% *************************************************************************
%
%\noindent\textbf{Algorithm Target: The Optimal LQR Controller.}
%
Under the above assumptions, the LQR optimal control $u^{*}$ associated with the quadruple $(A, B, Q, R)$ exists, is unique, and assumes the form of a full-state feedback control law \cite{AAR_multivariable:book}
\begin{align}
	u^{*}
	=
	- K^{*} x,
    \label{eq:Kstar_LQR}
\end{align}

\noindent where $K^{*} = R^{-1} B^{T} P^{*}$, and $P^{*} \in \mathbb{S}^{n}$, $P^{*} > 0$ is the unique positive definite solution to the control algebraic Riccati equation (CARE)
\begin{align}
	A^{T} P^{*} + P^{*} A - P^{*} B R^{-1} B^{T} P^{*} + Q
	=
	0.
	\label{eq:CARE}
\end{align}

% *************************************************************************
%
% KLEINMAN'S
%
% *************************************************************************

\noindent\textbf{Kleinman's Algorithm \cite{Kleinman_AREs:1968}.}
Suppose that $K_{0} \in \mathbb{R}^{m \times n}$ is such that $A - B K_{0}$ is Hurwitz. For iteration $i$ $(i = 0, 1, \dots)$, let $P_{i} \in \mathbb{S}^{n}$, $P_{i} > 0$ be the symmetric positive definite solution of the ALE
\begin{align}
	(A - B K_{i})^{T} P_{i} + P_{i} (A - B K_{i}) + K_{i}^{T} R K_{i} + Q
	=
	0.
	\label{eq:Kleinman_LE}
\end{align}
\noindent Having solved the ALE $P_{i}$ (\ref{eq:Kleinman_LE}), the controller $K_{i+1} \in \mathbb{R}^{m \times n}$ is updated recursively as
\begin{align}
	K_{i+1}
	=
	R^{-1} B^{T} P_{i}.
	\label{eq:Kleinman_controller_update}
\end{align}

% *************************************************************************
%
% THEOREM: KLEINMAN CARE
%
% *************************************************************************

\begin{theorem}[Stability, Convergence of Kleinman's Algorithm \cite{Kleinman_AREs:1968}]\label{thm:Kleinman_AREs}
Let the preceding assumptions of this section hold. Then we have the following:

\begin{enumerate}[(i)]

	\item $A - B K_{i}$ is Hurwitz for all $i \geq 0$.
	
	\item $P^{*} \leq P_{i+1} \leq P_{i}$ for all $i \geq 0$.
	
	\item $\lim\limits_{i \rightarrow \infty} K_{i} = K^{*}$, $\lim\limits_{i \rightarrow \infty} P_{i} = P^{*}$.
	
\end{enumerate}

\end{theorem}

%\FloatBarrier

%\fi				% Comment this line and \iffalse to include section

% *************************************************************************
%
% ALGORITHMS AND TRAINING
%
% *************************************************************************

%\iffalse			% Comment this line and \fi to include section

% ************************************************************************
% ************************************************************************
% ************************************************************************
%
% ALGORITHMS AND TRAINING
%
% ************************************************************************
% ************************************************************************
% ************************************************************************

% ***********************
%
% RELATIVE PATH TO FIGURES
%
\renewcommand{\relpath}{figures/algs_training/}

\subsection{Decentralized Excitable Integral Reinforcement Learning (dEIRL)}\label{sec:dEIRL}

The original EIRL work \cite{\TNNLSdEIRLCitation} develops a suite of learning algorithms. In this section, we focus on the flagship decentralized method: dEIRL, but note that the results here apply to the full suite just as readily. 
Inspired by Kleinman's approach, dEIRL iteratively solves the CARE associated with the linearization of the \emph{nonlinear} system (\ref{eq:sys_nonlin_2x2}) via a sequence of simpler linear regression problems, reducing the dimension of the regressions by taking advantage of the decentralized dynamical structure (\ref{eq:sys_nonlin_2x2}). 
In order to solve these regressions, dEIRL uses state-action data $(x_{j}, u_{j})$ generated in each decentralized loop $1 \leq j \leq N$ under the initial stabilizing controller $K_{0}$, collecting $l$ data samples at the sample period $T_{s}$. This data forms a learning regression related to the Kleinman's ALE which is solved for $i = i^{*}$ iterations to produce the final controller \cite{\TNNLSdEIRLCitation}. 
%
%We offer a full derivation of the dEIRL algorithm in the original development work \cite{\TNNLSdEIRLCitation}, so we keep the formulation brief here and focus on notational changes brought about by the new symmetric Kronecker product results of Section \ref{sec:skron}.

% *************************************************************************
%
% RELEVANT OPERATORS
%
% *************************************************************************

\noindent\textbf{Operators.}
The following maps are necessary for this study.

% *************************************************************************
%
% DEFINITION: RELEVANT OPERATORS
%
% *************************************************************************

\begin{definition}
For $l \in \mathbb{N}$ and a strictly increasing sequence $\{t_{k}\}_{k=0}^{l}$, whenever $x, y : [t_{0}, t_{l}] \rightarrow \mathbb{R}^{n}$, define the matrix $\delta_{x, y} \in \mathbb{R}^{l \times \underline{n}}$ as
\begin{align}
	\delta_{x,y}
	&=
	\left[
	\begin{array}{c}
	\big( x(t_{1}) + y(t_{0}) \big)^{T} \skron \big( x(t_{1}) - y(t_{0}) \big)^{T}
	\\
	\big( x(t_{2}) + y(t_{1}) \big)^{T} \skron \big( x(t_{2}) - y(t_{1}) \big)^{T}
	\\
	\vdots
	\\
	\big( x(t_{l}) + y(t_{l-1}) \big)^{T} \skron \big( x(t_{l}) - y(t_{l-1}) \big)^{T}
	\end{array}
	\right],
	\label{eq:delta_xy_def}
\end{align}

\noindent where $\skron$ denotes the symmetric Kronecker product (cf. Section \ref{sec:skron}).
Whenever $x, y$ are square-integrable, define $I_{x, y} \in \mathbb{R}^{l \times \underline{n}}$ as
\begin{align}
	I_{x, y}
	&=
	\left[
	\begin{array}{c}
	\int_{t_{0}}^{t_{1}} x^{T} \skron y^{T} \, d\tau
	\\
	\int_{t_{1}}^{t_{2}} x^{T} \skron y^{T} \, d\tau
	\\
	\vdots
	\\
	\int_{t_{l-1}}^{t_{l}} x^{T} \skron y^{T} \, d\tau
	\end{array}
	\right].	
	\label{eq:I_xy_def}
\end{align}

\end{definition}

% *************************************************************************
%
% ALGORITHM
%
% *************************************************************************

\noindent\textbf{Algorithm.}
In the following, let any loop $1 \leq j \leq N$ be given. Suppose that $K_{0,j} \in \mathbb{R}^{m_{j} \times n_{j}}$ is such that $A_{jj} - B_{jj} K_{0,j}$ is Hurwitz in loop $j$. 
\noindent Manipulating (\ref{eq:sys_nonlin_2x2}), we have
\begin{align}
	\dot{x}_{j}
	&=
	A_{i,j} x_{j} + B_{jj} K_{i,j} x_{j} + g_{j}(x) u + w_{j},
	\nonumber
	\\
	w_{j}
	&\triangleq
	f_{j}(x) - A_{jj} x_{j},
	\qquad
	A_{i,j}
	\triangleq
	A_{jj} - B_{jj} K_{i,j}.
	\label{eq:nonlin_sys_dirl_rewritten}
\end{align}

\noindent Given $l_{j} \in \mathbb{N}$ and a strictly increasing sequence $\{t_{k,j}\}_{k=0}^{l_{j}}$, following the derivation presented in \cite{\TNNLSdEIRLCitation}, we arrive at the least-squares regression
\begin{align}
	\mathbf{A}_{i,j} \svec(P_{i,j}) 
	&=
	\mathbf{b}_{i,j},
	\label{eq:DIRL_nonlin_lsq}
\end{align}

\noindent where $\svec$ denotes the symmetric vectorization operator (cf. Section \ref{sec:skron}), and $\mathbf{A}_{i,j} \in \mathbb{R}^{l_{j} \times \underline{n}_{j}}$, $\mathbf{b}_{i,j} \in \mathbb{R}^{l_{j}}$ are given by
\begin{align}
	\mathbf{A}_{i,j}
	&=
	\delta_{x_{j}, x_{j}} 
	\nonumber
	\\
	&\,
	- 2 \big[  I_{x_{j}, x_{j}} \left( I_{n_{j}} \skron B_{jj} K_{i,j} \right)^{T}
	+ I_{x_{j}, g_{j}u} + I_{x_{j}, w_{j}} \big],
	\label{eq:DIRL_nonlin_lsq_A}
	\\
	\mathbf{b}_{i,j}
	&=
	-I_{x_{j}, x_{j}} \svec (Q_{i,j}), 
	\;\;
    Q_{i,j} 
	\triangleq
	Q_{j} + K_{i,j}^{T} R_{j} K_{i,j}.
	\label{eq:DIRL_nonlin_lsq_b}
\end{align}

Having performed the regression $\svec(P_{i,j})$ (\ref{eq:DIRL_nonlin_lsq}), we update the controller analogously to (\ref{eq:Kleinman_controller_update}):
\begin{align}
    K_{i+1,j} 
    = 
    R_{j}^{-1} B_{jj}^{T} P_{i,j}.
    \label{eq:DIRL_controller_update}
\end{align}

% *************************************************************************
%
% THEOREM: EQUIVALENCE OF IRL AND KLEINMAN RECURSION
%
% *************************************************************************

We conclude with the key stability/convergence result of the dEIRL algorithm, proven in \cite{\TNNLSdEIRLCitation}:

\begin{theorem}[Equivalence of dEIRL Algorithm and Kleinman's Algorithm \cite{\TNNLSdEIRLCitation}]\label{thm:DIRL_nonlin_Kleinman_equivalence}
Suppose for $1 \leq j \leq N$ that $l_{j} \in \mathbb{N}$ and the sample instants $\{t_{k,j}\}_{k=0}^{l_{j}}$ are chosen such that the matrix $I_{x_{j},x_{j}}$ (\ref{eq:I_xy_def}) has full column rank $\underline{n}_{j}$. If $K_{0,j}$ is such that $A_{jj} - B_{jj} K_{0,j}$ is Hurwitz, then the dEIRL algorithm and Kleinman's algorithm are equivalent in that the sequences $\{P_{i,j}\}_{i=0}^{\infty}$ and $\{K_{i,j}\}_{i=0}^{\infty}$ produced by both are identical. Thus, the stability/convergence conclusions of Theorem \ref{thm:Kleinman_AREs} hold for the dEIRL algorithm as well.

\end{theorem}

%\FloatBarrier

%\fi				% Comment this line and \iffalse to include section

% *************************************************************************
%
% MOTIVATING EXAMPLE
%
% *************************************************************************

%\iffalse			% Comment this line and \fi to include section

% ************************************************************************
% ************************************************************************
% ************************************************************************
%
% MOTIVATING EXAMPLE
%
% ************************************************************************
% ************************************************************************
% ************************************************************************

\section{A Motivating Example}\label{sec:motivating_ex}

% ************************************************************************
% ************************************************************************
% ************************************************************************
%
% SUBSECTION: EVALUATION STUDIES -- NOMINAL MODEL
%
% ************************************************************************
% ************************************************************************
% ************************************************************************

% ***********************
%
% RELATIVE PATH TO FIGURES
%
\renewcommand{\relpath}{figures/}

In this section, we motivate the need for the developed modulation-enhanced excitation (MEE) framework via an illustrative example. Consider the system
\begin{align}
	\left[
	\begin{array}{c}
		\dot{x}_{1}
		\\
		\dot{x}_{2}
	\end{array}
	\right]
	=
	\left[
	\begin{array}{cc}
		-1 & 0
		\\
		0 & -0.1
	\end{array}
	\right]	
	\left[
	\begin{array}{c}
		x_{1}
		\\
		x_{2}
	\end{array}
	\right]
	+
	\left[
	\begin{array}{cc}
		1 & 0
		\\
		0 & 1
	\end{array}
	\right]	
	\left[
	\begin{array}{c}
		u_{1}
		\\
		u_{2}
	\end{array}
	\right].
	\label{eq:sys_lin2d}
\end{align}
(\ref{eq:sys_lin2d}) is a diagonal linear system consisting of a high-bandwidth loop $j = 1$ (associated with $x_{1}, u_{1}$) and a low-bandwidth loop $j = 2$ (associated with $x_{2}, u_{2}$). Such weakly-coupled two-loop systems with a decade separation in bandwidth are quite common in real-world applications (see, e.g., the HSV example of Section \ref{ssec:ES_hsv}). 

Oftentimes, control/actuator saturation is a major concern for designing in hardware, especially for RL excitation purposes in lower-bandwidth loops. For example, in robotics applications, designers must consider the load specifications of the motors/servomechanisms, which have significant impact on achievable closed-loop performance \cite{Mondal_AA_Rodriguez_Manne_Das_BA_Wallace_wheeled_robot:2019,Mondal_BA_Wallace_AA_Rodriguez_wheeled_robot:2020}. In aerospace applications, control surface deflections have strict bounds to avoid aerodynamic stall \cite{Stengel_flight_dynamics:book:2022}. To account for control saturation concerns in this example, suppose the designer has constraints of $u_{1}(t) \in [-1, 1]$ in the high-bandwidth loop $j = 1$ and $u_{2}(t) \in [-0.1, 0.1]$ in the low-bandwidth loop $j = 2$. We consider natural designer first-choices for the cost structure: $Q = I_{2}$, $R = \texttt{diag}(1, 10)$, where the larger control penalty on $u_{2}$ reflects the designer's control saturation concerns in this channel. Similarly, for EIRL's hyperparameters, we select a sample period $T_{s} = 0.1$ (i.e., a sample frequency approximately a decade above the highest-bandwidth mode in the plant), $i^{*} = 5$ iterations, and $l = 5$ data points. For probing noises, we choose $d_{1}(t) = \cos(t)$, $d_{2}(t) = 0.1 \cos(0.1t)$, the amplitudes reflecting the designer's consideration of control saturation in each channel, and the frequencies placed at the bandwidths of the respective loop modes.

Suppose that the designer opts for a single-loop design $N = 1$ (i.e., not using decentralization and hence executing EIRL rather than dEIRL \cite{\TNNLSdEIRLCitation}). Noting that the open-loop system (\ref{eq:sys_lin2d}) is stable, Theorem \ref{thm:DIRL_nonlin_Kleinman_equivalence} guarantees that the stabilizing controller $K_{0} = 0 \in \mathbb{R}^{2 \times 2}$ will result in convergence of the (d)EIRL algorithm to the optimal controller $K^{*}$. 
Running EIRL, the final controller $K_{i^{*}}$ converges to within $1.62 \times 10^{-9}$ of the optimal $K^{*}$. However, the learning regression has a large peak condition number of 138.47 (cf. Section \ref{ssec:ES_lin2d} for complete evaluation data). Intuitively, the culprit stems from the max control effort requirements placed in the low-bandwidth loop $j = 2$. As a result, the designer can excite the low-bandwidth loop $j = 2$ at only a tenth the control effort of the high-bandwidth loop $j = 1$. Since the probing noise frequencies were placed at the respective loop bandwidths, the state response $x_{1}(t)$ in the high-bandwidth loop exhibits approximately ten times the amplitude of the response $x_{2}(t)$ in the low-bandwidth loop, resulting in scaling and thus conditioning issues in the regression matrix $\mathbf{A}_{i,j}$ (\ref{eq:DIRL_nonlin_lsq_A}). 

The designer's insight to fix the issue is clear: The state response $x_{2}(t)$ in the low-bandwidth loop needs to be scaled up by a factor of ten to improve scaling. 
This raises the central questions: How may we address this significant scaling issue in a systematic design framework which leverages physical insights (in this case, our saturation constraints) while achieving excitation and thus good numerical conditioning? Crucially, how can we ensure that such a framework preserves dEIRL's key theoretical convergence and closed-loop stability guarantees? 
As will be shown below, MEE is the answer to these questions. First, however, we must develop some essential symmetric Kronecker product results.

In a real-world analogue to this scenario, the designer oftentimes has no physical means of recourse to address these conditioning issues: The excitation level in the high-bandwidth loop $j = 1$ cannot be reduced without degrading PE and hence learning performance in this loop. Furthermore, oftentimes unit scaling between unlike physical measurements renders the equilibration of responses physically intractable (e.g., in the HSV example studied in Section \ref{ssec:ES_hsv}, velocity oscillations on the order of 100 ft/s are needed to achieve good PE in the translational loop, yet flightpath angle oscillations on the order of 100 deg in the rotational loop are nonsensical). 
%Prior ADP-based CT-RL algorithms do not address this subtlety of scaling, leaving the designer only a probing noise excitation to achieve PE and thereby modulate the numerical quality of the underlying learning regression \cite{BA_Wallace_J_Si_CT_RL_review:2022}. 
This simple example illustrates that the problem runs deeper: Even when the system has been excited to the greatest possible extent, physical constraints and/or unit intermingling may still leave the learning regression poorly conditioned. These fundamental design concerns make the symmetric Kronecker product results of the next section all the more vital.

%\FloatBarrier

%\fi				% Comment this line and \iffalse to include section

% *************************************************************************
%
% SYMMETRIC KRONECKER PRODUCT
%
% *************************************************************************

%\iffalse			% Comment this line and \fi to include section

% ************************************************************************
% ************************************************************************
% ************************************************************************
%
% SECTION: SYMMETRIC KRONECKER PRODUCT
%
% ************************************************************************
% ************************************************************************
% ************************************************************************

% ************************************************************************
%
% FIGURE: (r, c) ROW/COLUMN INDEXING OPERATIONS
%
% ************************************************************************

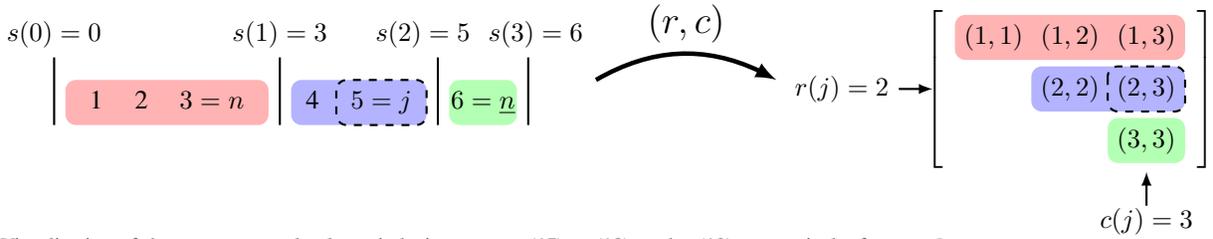
\begin{figure*}[ht]
\begin{center}
\begin{tikzpicture}[scale=0.6]
	% s(0)
	\draw [thick] (0,0)
		-- (0,1.5)
		node [above] {$s(0) = 0$}
		coordinate (tmp)
		;		% END
	\coordinate (tmp) at (tmp |- 0,0);
	% First rectangle	  
	\fill[fill=red!30, rounded corners] ([shift={(0.25, 0)}]tmp) 
	  	rectangle node {
	  	\begin{tabular}{ccc}
	  	1 & 2 & $3 = n$
	  	\end{tabular}
	  	}
	  	++(4.5,1)
	  	coordinate (tmp)
	  	;		% END
	\coordinate (tmp) at (tmp |- 0,0);
	% s(1)
	\draw [thick] ([shift={(0.25, 0)}]tmp) 
		-- ++(0,1.5)
		node [above] {$s(1) = 3$}
		coordinate (tmp)
		;		% END
	\coordinate (tmp) at (tmp |- 0,0);
	% Second rectangle
  	\fill[fill=blue!30, rounded corners] ([shift={(0.25, 0)}]tmp)  
	  	rectangle node {
	  	\begin{tabular}{cc}
	  	4 & $5 = j$
	  	\end{tabular}
	  	}
	  	++(3,1)
		coordinate (tmp)
		;		% END
	\coordinate (tmp) at (tmp |- 0,0);
	% Box around j
  	\draw[dashed, thick, rounded corners] ([shift={(-2.0, 1.0)}]tmp) rectangle (tmp);
	% s(2)
	\draw [thick] ([shift={(0.25, 0)}]tmp) 
		-- ++(0,1.5)
		node [above, shift={(-0.2,0)}] {$s(2) = 5$}
		coordinate (tmp)
		;		% END	
 	\coordinate (tmp) at (tmp |- 0,0);
 	% Third rectangle
 	\fill[fill=green!30, rounded corners] ([shift={(0.25, 0)}]tmp)  
	  	rectangle node {
	  	\begin{tabular}{c}
	  	$6 = \underline{n}$
	  	\end{tabular}
	  	}
	  	++(1.5,1)
		coordinate (tmp)
		;		% END	 
	\coordinate (tmp) at (tmp |- 0,0);
	% s(3)
	\draw [thick] ([shift={(0.25, 0)}]tmp) 
		-- ++(0,1.5)
		node [above, shift={(0.1,0)}] {$s(3) = 6$}
		coordinate (tmp)
		;		% END	 	
	\coordinate (tmp) at (tmp |- 0,0);
	% Mapping arrow
	\draw [-\Arrowbf, line width = \Linewidthbf, bend left] ([shift={(1.5,1)}]tmp)
		to [bend left]
		node [pos=0.5, above] {\Large $(r, c)$}
		++(4,0)
		coordinate (tmp)
		;		% END
	\coordinate (tmp) at (tmp |- 0,0);
	% Matrix
	\matrix (M1) [matrix of nodes,{left delimiter=[},{right delimiter=]}, row sep=0.1cm] at ([shift={(6.5,0.8)}]tmp)
		{
			$(1,1)$ & $(1,2)$ & $(1,3)$ 
			\\
			 & $(2,2)$ & $(2,3)$   
			\\
			 &  & $(3,3)$ 
			\\
		};		% END
	% Colored rectangles inside matrix
	\begin{pgfonlayer}{background}
		% First rectangle
		\fill[fill=red!30, rounded corners] (M1-1-1.north west) rectangle (M1-1-3.south east);	
		% Second rectangle
		\fill[fill=blue!30, rounded corners] (M1-2-2.north west) rectangle (M1-2-3.south east);		
		% Third rectangle
		\fill[fill=green!30, rounded corners] (M1-3-3.north west) rectangle (M1-3-3.south east);			  	
	\end{pgfonlayer}
	% Box around (r(j), c(j))
  	\draw[dashed, thick, rounded corners] (M1-2-3.north west) rectangle (M1-2-3.south east); 	  
	% r(j) label
	\coordinate (tmp) at ([shift={(-0.3,0)}]M1.west |- M1-2-3);
	\draw [latex-, thick] (tmp)
		-- ([shift={(-0.75,0)}]tmp)
		node [left, shift={(0,0)}] {$r(j) = 2$}
		;		% END
	% c(j) label
	\coordinate (tmp) at (M1-2-3 |- M1.south);
	\draw [latex-, thick] (tmp)
		-- ([shift={(0,-0.75)}]tmp)
		node [below, shift={(0,0.1)}] {$c(j) = 3$}
		;		% END		
\end{tikzpicture}
\vspace*{-0.2in}
\caption{Visualization of the sum, row, and column indexing maps $s$ (\ref{eq:sj_def}), $r$ (\ref{eq:rj_def}), and $c$ (\ref{eq:cj_def}), respectively, for $n = 3$.}
\label{fig:rj_cj}
\end{center}
\end{figure*}

% ************************************************************************
%
% BEGIN SECTION
%
% ************************************************************************

\section{The Symmetric Kronecker Product \& Symmetric Kronecker Sum}\label{sec:skron}

In this section, we first provide an overview of the symmetric Kronecker product, summarizing the notable developments to-date. We then derive a construction of the map and prove new key properties necessary for the development of the proposed MEE framework.

% ************************************************************************
% ************************************************************************
% ************************************************************************
%
% SUBSECTION: CONSTRUCTION
%
% ************************************************************************
% ************************************************************************
% ************************************************************************

\subsection{Overview}\label{ssec:skron_overview}

The symmetric Kronecker product was originally devised in \cite{Alizadeh_Haeberly_Overton_skron:1998} for application to semidefinite programming as an operation on square-symmetric matrices. In this context, it was shown that the symmetric Kronecker product $\skron$ is symmetric as a bilinear form: $A \skron B = B \skron A$, and that $(A \skron A)^{-1} = A^{-1} \skron A^{-1}$ in the case $A$ is invertible. The spectrum of $A \skron B$ was identified in the case that $A, B$ are symmetric and commute. 
The symmetric Kronecker product was then extended in \cite{Todd_Toh_Tutuncu_skron:1998} to an operation on arbitrary square matrices. \cite{Todd_Toh_Tutuncu_skron:1998} identified many key algebraic properties analogous to those of the standard Kronecker product, including the usual transposition, mixed product, and mixed vector product identities. The spectrum of $A \skron A$ was identified in the general square matrix case. 
\cite{Tuncel_Wolkowicz_skron:2005} then identified eigenpair relationships and definiteness characterizations: that positive (semi)definiteness of $A \skron B$ is equivalent to that of $A \otimes B$.
More recently, the works \cite{Kalantarova_Tuncel_skron:2021,Kalantarova_PhD_thesis_skron:2019} provide spectral interlacing properties of the related Jordan-Kronecker product.

Notably, prior works to date have treated the symmetric Kronecker product as an operation only on square matrices $A, B \in \mathbb{R}^{n \times n}$, which we here generalize to rectangular matrices $A, B \in \mathbb{R}^{m \times n}$.
%, including the vector case $n = 1$. 
Among other advantages, this allows us to identify the eigenstructure of $A \skron B$ as relating to the symmetric Kronecker products $x \skron y$ of eigenvectors $x, y$ of $A$ and $B$ -- a critical parallel to the well-known result of the standard Kronecker product. 
% We generalize the symmetric Kronecker product algebraic identities to the rectangular-matrix case, and 
We also prove new properties in the square case which will be essential to the development of MEE. Importantly, we introduce the concept of the symmetric Kronecker sum $\skrons$, proving algebraic, spectral, and exponentiation properties, as well as its role in characteriing existence/uniqueness of solutions to ALEs.
%in the symmetric matrix case, a role well-understood of its standard counterpart $\oplus$. 

% ************************************************************************
% ************************************************************************
% ************************************************************************
%
% SUBSECTION: CONSTRUCTION
%
% ************************************************************************
% ************************************************************************
% ************************************************************************

\subsection{Construction}\label{ssec:skron_construction}

Prior formulations of the symmetric Kronecker product \cite{Alizadeh_Haeberly_Overton_skron:1998,Todd_Toh_Tutuncu_skron:1998,Tuncel_Wolkowicz_skron:2005,Kalantarova_Tuncel_skron:2021,Kalantarova_PhD_thesis_skron:2019} first define the product implicitly, but here we move straight to an explicit construction.
For $n \in \mathbb{N}$, let $\{E_{i}\}_{i=1}^{\underline{n}}$ denote the orthonormal basis for $\left( \mathbb{S}^{n}, \left< \cdot, \cdot \right>_{F} \right)$ enumerated as follows. Define $s : \{0, \dots, n\} \rightarrow \{0, \dots, \underline{n}\}$, $r, c : \{1, \dots, \underline{n}\} \rightarrow \{1, \dots, n\}$ by
\begin{align}
	s(j)
	&=
	\sum_{i=1}^{j} (n - (i-1)),
	\label{eq:sj_def}
	\\
	r(j)
	&=
	p, \qquad s(p-1) < j \leq s(p),
	\label{eq:rj_def}
	\\
	c(j)
	&=
	(r(j) - 1) + \left(j - s \big( r(j) - 1 \big) \right).
	\label{eq:cj_def}
\end{align}
\noindent When necessary, we will add subscripts $s_{n}$, $r_{n}$, $c_{n}$ to these maps to make their associated dimension $n$ explicit.
Note that $\{(r(j), c(j))\}_{j=1}^{\underline{n}}$ is given by $(1,1), (1,2)$, $\dots$, $(1,n)$, $(2,2), (2,3)$, $\dots, (2, n)$, $(3,3), \dots, (n-1,n), (n,n)$. This associates the index $1 \leq j \leq \underline{n}$ with its corresponding 
\linebreak 
row/column index $(r(j), c(j))$ on/above the diagonal, beginning at the first row/column and moving left to right, up to down (cf. Figure \ref{fig:rj_cj}). These maps have not been defined explicitly in the constructions of prior works \cite{Alizadeh_Haeberly_Overton_skron:1998,Todd_Toh_Tutuncu_skron:1998,Tuncel_Wolkowicz_skron:2005,Kalantarova_Tuncel_skron:2021,Kalantarova_PhD_thesis_skron:2019}; however, subsequently they will show great utility in indexing operations for proving properties of the symmetric Kronecker product, especially in developing our results for the rectangular-matrix case.
Letting $\{e_{i}\}_{i=1}^{n}$ denote the standard basis on $\mathbb{R}^{n}$, we are now ready to enumerate the orthonormal basis $\{E_{j}\}_{j=1}^{\underline{n}}$ as
\begin{align}
	E_{j}
	=
	\begin{cases}
		e_{r(j)} e_{c(j)}^{T}, & r(j) = c(j),
		\\
		\frac{\sqrt{2}}{2} \left( e_{r(j)} e_{c(j)}^{T} + e_{c(j)} e_{r(j)}^{T} \right), & r(j) < c(j).
	\end{cases}
	\label{eq:Sn_ONB}
\end{align}
\noindent Define $W \in \mathbb{R}^{\underline{n} \times n^{2}}$ as 
\begin{align}
	W
	=
	\left[ 
	\begin{array}{c}
		\textnormal{vec}^{T}(E_{1})
		\\
		\vdots
		\\
		\textnormal{vec}^{T}(E_{\underline{n}})
	\end{array}
	\right].
	\label{eq:W_def}
\end{align}
\noindent Whenever necessary, we will also add a subscript $W_{n} \in \mathbb{R}^{\underline{n} \times n^{2}}$ to this matrix to make its dimensions explicit. 
%
% *************************************************************************
%
% DEFINITION: SYMMETRIC VECTORIZATION, PROJECTION
%
% *************************************************************************
%
\begin{definition}[Symmetric Vectorization, Orthogonal Projection]
Define $\svec : \mathbb{S}^{n} \rightarrow \mathbb{R}^{\underline{n}}$ and $\pi: \mathbb{R}^{n \times n} \rightarrow \mathbb{S}^{n}$ by
\begin{align}
	\svec(P)
	&=
	\big[
	p_{1,1}, \,  \sqrt{2} p_{1,2}, \dots, \,  \sqrt{2} p_{1,n}, 
	\nonumber
	\\
	&\qquad\qquad
	p_{2,2}, \,  \sqrt{2} p_{2,3}, \dots, \,  \sqrt{2} p_{n-1,n}, \, p_{n,n}
	\big]^{T}
	\nonumber
	\\
	&=
	\big[ \left< P, E_{1} \right>_{F}, \dots, \left< P, E_{\underline{n}} \right>_{F} \big]^{T},
	\label{eq:svec_def}		
	\\
	\pi(A)
	&=
	\frac{A + A^{T}}{2},
	\label{eq:pi_def}	
\end{align}
\noindent and define $\smat = \svec^{-1} : \mathbb{R}^{\underline{n}} \rightarrow \mathbb{S}^{n}$. We will discuss the properties of these operators shortly (cf. Proposition \ref{prop:skron_HS_iso}).

\end{definition}

%
% *************************************************************************
%
% DEFINITION: SYMMETRIC KRONECKER PRODUCT
%
% *************************************************************************
%
\begin{definition}[The Symmetric Kronecker Product]
Define the symmetric Kronecker product $\skron : \mathbb{R}^{m \times n} \times \mathbb{R}^{m \times n} \rightarrow \mathbb{R}^{\underline{m} \times \underline{n}}$ as
\begin{align}
	A \skron B
	=
	W_{m} \left( A \otimes B \right) W_{n}^{T}.
\end{align}
\end{definition}
%
% *************************************************************************
%
% DEFINITION: SYMMETRIC KRONECKER SUM
%
% *************************************************************************
%
\begin{definition}[The Symmetric Kronecker Sum]
Define the symmetric Kronecker sum $\skrons : \mathbb{R}^{n \times n} \times \mathbb{R}^{n \times n} \rightarrow \mathbb{R}^{\underline{n} \times \underline{n}}$ as
\begin{align}
	A \skrons B
	=
	A \skron I + I \skron B
	=
	(A + B) \skron I.
\end{align}
\end{definition}

% ************************************************************************
% ************************************************************************
% ************************************************************************
%
% SUBSECTION: PROPERTIES
%
% ************************************************************************
% ************************************************************************
% ************************************************************************

\subsection{Properties}\label{ssec:skron_propts}

%In this section, we establish the key properties of the symmetric Kronecker product and symmetric Kronecker sum which we employ in development the proposed dEIRL prescaling framework in Section \ref{sec:prescaling}. 
We begin this section by outlining the interaction of the vectorization operations $\vecc$, $\svec$ with the Frobenius inner product on matrix spaces.

% *************************************************************************
%
% PROPOSITION: KRONECKER PRODUCT PROPERTIES
%
% *************************************************************************

\begin{proposition}[Vectorization and Frobenius Hilbert Space Structure]\label{prop:skron_HS_iso}
\begin{enumerate}[1)]
	\item $\vecc : \left( \mathbb{R}^{m \times n}, \left< \cdot, \cdot \right>_{F} \right) \rightarrow \left( \mathbb{R}^{mn}, \left< \cdot, \cdot \right> \right)$ is a Hilbert space isomorphism; i.e., a linear bijection for which $\vecc^{T}(A) \vecc(B) = \left< A, B \right>_{F}$, $A, B \in \mathbb{R}^{m \times n}$.
	
\end{enumerate}

\begin{enumerate}[1S)]

	\item $\svec : \left( \mathbb{S}^{n}, \left< \cdot, \cdot \right>_{F} \right) \rightarrow \left( \mathbb{R}^{\underline{n}}, \left< \cdot, \cdot \right> \right)$ is a Hilbert space isomorphism; i.e., a linear bijection for which $\svec^{T}(A) \svec(B) = \left< A, B \right>_{F}$, $A, B \in \mathbb{S}^{n}$.
	
\end{enumerate}

\begin{enumerate}[1)]

	% Start at 2
	\setcounter{enumi}{1}

	\item In the square-matrix case, the operators $\vecc, \svec$ interact with the Hilbert space structure of $\left( \mathbb{R}^{n \times n}, \left< \cdot, \cdot \right>_{F} \right)$ via the following commutative diagram:
	\begin{equation}
	\begin{tikzcd}
	%\begin{tikzcd}[cells={nodes={draw=gray}}]
		\left( \mathbb{R}^{n \times n}, \left< \cdot, \cdot \right>_{F} \right) \arrow[r, "\pi", "\perp \textnormal{ proj.}"'] \arrow[d, xshift=-1.2ex, "\cong", "\vecc"']
		&
		\left( \mathbb{S}^{n}, \left< \cdot, \cdot \right>_{F} \right) \arrow[d, xshift=-1.2ex, "\cong", "\vecc"']  \arrow[dr, "\cong" yshift=-0.8ex, xshift=0.7ex,  "\svec"']
		&
		\\
		\left( \mathbb{R}^{n^{2}}, \left< \cdot, \cdot \right> \right) \arrow[r, "W^{T} W", "\perp \textnormal{ proj.}"'] \arrow[u, xshift=1.2ex, "\mat"']
		&
		\left( \vecc(\mathbb{S}^{n}), \left< \cdot, \cdot \right> \right) \arrow[r, yshift=0.9ex, "W", "\cong"'] \arrow[u, xshift=1.2ex, "\mat"']  
		&
		\left( \mathbb{R}^{\underline{n}}, \left< \cdot, \cdot \right> \right) \arrow[l, yshift=-0.9ex, "W^{T}"] \arrow[ul, end anchor={[xshift=-4ex,yshift=1ex]south east}, xshift=4.5ex,  "\smat"' ]
	\end{tikzcd}
	\label{diag:svec}
	\end{equation}
	\noindent where $W \in \mathbb{R}^{\underline{n} \times n^{2}}$ is given by (\ref{eq:W_def}), and $\cong$ denotes Hilbert space isomorphism.
	
\end{enumerate}

\end{proposition}

In (\ref{diag:svec}), $\vecc(\mathbb{S}^{n}) \subset \mathbb{R}^{n^{2}}$ is often called the space of symmetric vectors \cite{Kalantarova_Tuncel_skron:2021,Kalantarova_PhD_thesis_skron:2019}. The operator $\pi$ (\ref{eq:pi_def}) and the matrix $W^{T} W \in \mathbb{R}^{n^{2} \times n^{2}}$ are the orthogonal projections onto the symmetric matrices and symmetric vectors, respectively. Furthermore, (\ref{diag:svec}) shows that the order of vectorization and projection may be swapped by interchanging these two projections.

We recall the general result from linear algebra that, given an $n$-dimensional real vector space $V$, any basis $\{x_{i}\}_{i=1}^{n}$ for $V$ establishes a linear isomorphism $V \leftrightarrow \mathbb{R}^{n}$ between Euclidean space and elements of $V$ via their (unique) representation in the basis. In the case considered, the symmetric vectors $\vecc(\mathbb{S}^{n}) \subset \mathbb{R}^{n^{2}}$ comprise an $\underline{n}$-dimensional subspace spanned by the (orthonormal) basis $\{ \vecc(E_{i}) \}_{i = 1}^{\underline{n}}$ (\ref{eq:Sn_ONB}). $W \in \mathbb{R}^{n^{2} \times \underline{n}}$ is the matrix representation of the linear isomorphism $\vecc(\mathbb{S}^{n}) \leftrightarrow \mathbb{R}^{\underline{n}}$. Indeed, invoking the Hilbert space structure of $\vecc(\mathbb{S}^{n})$, orthonormality of $\{ \vecc(E_{i}) \}_{i = 1}^{\underline{n}}$ implies the $i$-th coefficient representation $(i = 1, \dots, \underline{n})$ of an element $\vecc(P) \in \vecc(\mathbb{S}^{n})$ is given by $\left< \vecc(P) , \vecc(E_{i}) \right> = \left< P, E_{i} \right>_{F}$, which is precisely the action of $W$ (\ref{eq:W_def}). Similarly, examination of (\ref{eq:svec_def}) immediately shows that the symmetric vectorization $\svec$ is the Euclidean space correspondence $\mathbb{S}^{n} \leftrightarrow \mathbb{R}^{\underline{n}}$ associated with the orthonormal basis $\{E_{i}\}_{i=1}^{\underline{n}}$.
In all, (\ref{diag:svec}) illustrates that $\svec = W \circ \vecc$, a composition of Hilbert space isomorphisms, is itself a Hilbert space isomorphism.
Meanwhile, $\vecc$ viewed as a map $\mathbb{S}^{n} \rightarrow \mathbb{R}^{n^{2}}$ ``vectorizes" elements of $\mathbb{S}^{n}$ into the Euclidean space $\mathbb{R}^{n^{2}}$ isometrically, but $\vecc$ is not onto. Thus, $\svec$ is the natural Hilbert space isomorphism of study on the symmetric matrices, as $\vecc$ is to $\mathbb{R}^{n \times n}$.

Having formally related the properties of the vectorization operations $\vecc$ and $\svec$, it is perhaps of no surprise that $\svec$ algebraically interacts with the symmetric Kronecker product $\skron$ in an entirely analogous fashion to the interaction between $\vecc$ and the Kronecker product $\otimes$. 
%We formalize these notions below, first providing well-known properties of the Kronecker product in Proposition \ref{prop:kron_propts} and then their analogues for the symmetric Kronecker product in Proposition \ref{prop:skron_propts}. As a note, we have enumerated matrix dimensions in the hypotheses in order to make the relations as explicit as possible. Although this practice is commonly omitted for brevity, it is further necessary here now that we treat the rectangular-matrix case for the first time.

% *************************************************************************
%
% PROPOSITION: KRONECKER PRODUCT PROPERTIES
%
% *************************************************************************

\begin{proposition}[Kronecker Product Properties]\label{prop:kron_propts}
For the sake of completeness and for comparison to the newly-developed results for the symmetric Kronecker product, we list the following well-known properties of the standard Kronecker product:

\begin{enumerate}[1)]

	\item\label{prop:kron_propts_bilin} $\otimes : \mathbb{R}^{m \times n} \times \mathbb{R}^{p \times q} \rightarrow \mathbb{R}^{mp \times nq}$ is bilinear.
	
	\item\label{prop:kron_propts_sym} $\otimes$ is \emph{not} symmetric; i.e., $A \otimes B \neq B \otimes A$, in general.
	
	\item\label{prop:kron_propts_mixed_vec_prod} $(A \otimes B) \vecc(C) = \vecc(B C A^{T})$, $A \in \mathbb{R}^{m \times n}$, $B \in \mathbb{R}^{p \times q}$, $C \in \mathbb{R}^{q \times n}$.
	
	\item\label{prop:kron_propts_transpose} $(A \otimes B)^{T} = A^{T} \otimes B^{T}$, $A \in \mathbb{R}^{m \times n}$, $B \in \mathbb{R}^{p \times q}$.
	
	\item\label{prop:kron_propts_inv} $(A \otimes B)^{-1} = A^{-1} \otimes B^{-1}$, $A \in \GL(n)$, $B \in \GL(m)$.
	
	\item\label{prop:kron_propts_mixed_prod} $(A \otimes B)(C \otimes D) = AC \otimes BD$, $A \in \mathbb{R}^{m \times n}$, $B \in \mathbb{R}^{p \times q}$, $C \in \mathbb{R}^{n \times r}$, $D \in \mathbb{R}^{q \times s}$.
	
	\item\label{prop:kron_propts_eig} For square matrices $A \in \mathbb{R}^{n \times n}$ and $B \in \mathbb{R}^{m \times m}$, if $\sigma(A) = \{\lambda_{i} \mid i = 1, \dots, n\}$ and $\sigma(B) = \{\mu_{j} \mid j = 1, \dots, m\}$, then $\sigma(A \otimes B) = \{ \lambda_{i} \mu_{j} \mid i = 1, \dots, n, \, j = 1, \dots, m\}$. Furthermore, if $x_{i} \in \mathbb{C}^{n}$, $y_{j} \in \mathbb{C}^{m}$ are eigenvectors corresponding to the eigenvalues $\lambda_{i}$ of $A$ and $\mu_{j}$ of $B$, respectively, then $x_{i} \otimes y_{j}$ is an eigenvector corresponding to the eigenvalue $\lambda_{i} \mu_{j}$ of $A \otimes B$.
	
	\item\label{prop:kron_propts_ApI_sym} $A \otimes I$ is symmetric if and only if $A$ is, $A \in \mathbb{R}^{n \times n}$.

	\item\label{prop:kron_propts_sym_pd} If $A \in \mathbb{S}^{m}, B \in \mathbb{S}^{n}$ are symmetric positive definite, then so is $A \otimes B$.

    \item\label{prop:kron_propts_zero} $A \otimes B = 0$ if and only if at least one $A, B = 0$, $A \in \mathbb{R}^{m \times n}$, $B \in \mathbb{R}^{p \times q}$.
 
	\item\label{prop:kron_propts_det} $\det(A \otimes B) = \det(A)^{m} \det(B)^{n}$, $A \in \mathbb{R}^{m \times m}$, $B \in \mathbb{R}^{n \times n}$.
	
	\item\label{prop:kron_propts_diag} For $A \in \mathbb{R}^{m \times m}$, $B \in \mathbb{R}^{n \times n}$, if $A, B$ are diagonal, then $A \otimes B$ is diagonal. If $A, B \neq 0$ and $A \otimes B$ is diagonal, then $A, B$ are diagonal.
	
	\item\label{prop:kron_propts_ApB_I} For $A \in \mathbb{R}^{m \times m}$, $B \in \mathbb{R}^{n \times n}$, $A \otimes B = I_{mn}$ if and only if $A = \lambda I_{m}$, $B = \frac{1}{\lambda} I_{n}$ for some $\lambda \neq 0$.
	
	\item\label{prop:kron_propts_LG_homo} The map $\Phi : \GL(n) \rightarrow \GL^{+}(n^{2})$,
	\begin{align}
		\Phi(A) = A \otimes A,
		\qquad 
		A \in \GL(n),
		\label{eq:Phi_def}
	\end{align}
	is a Lie group homomorphism with $\ker \Phi = \{\pm I\}$. 
	%
 	%$\left. \Phi \right|_{\GL^{+}(n)} : \GL^{+}(n) \rightarrow \GL^{+}(n^{2})$ is a Lie group isomorphism onto its image. 
 	$\left. \Phi \right|_{\GL^{+}(n)} : \GL^{+}(n) \rightarrow \GL^{+}(n^{2})$ is an injective Lie group homomorphism if and only if $n$ is odd. 
	In the case $n$ is odd, $\Phi(\GL(n)) = \Phi(\GL^{+}(n)) \hookrightarrow \GL^{+}(n^{2})$ is connected. In the case $n$ is even, $\Phi(\GL(n))$ has two connected components $\Phi(\GL^{+}(n)), \Phi(\GL^{-}(n)) \hookrightarrow \GL^{+}(n^{2})$.

\end{enumerate}

\end{proposition}

% *************************************************************************
%
% PROOF
%
% *************************************************************************
%
\textit{Proof:} 
\ref{prop:kron_propts_bilin})--\ref{prop:kron_propts_ApB_I}) are standard results; see, e.g., \cite{Horn_matrix_analysis:book:1991,Brewer_Kronecker_products:1978}. Enumerating $A = \{a_{i,j}\}_{i,j=1}^{m}$, $B = \{b_{k,l}\}_{k,l=1}^{n}$, \ref{prop:kron_propts_diag}) and \ref{prop:kron_propts_ApB_I}) follow from the Kronecker product indexing identity:
\begin{align}
	&\left( A \otimes B \right)_{(i-1)n + k, (j-1) n + l} 
	=
	a_{i,j} b_{k, l},
	\nonumber
	\\
	&
	\qquad
	i,j 
	=
	1, \dots, m,
	\quad
	k, l
	=
	1, \dots, n.
	\label{eq:kron_index_id}
\end{align}
For \ref{prop:kron_propts_LG_homo}), that $\Phi$ is a group homomorphism follows from  \ref{prop:kron_propts_mixed_prod}), and that $\ker \Phi = \{\pm I\}$ follows from \ref{prop:kron_propts_ApB_I}). For smoothness, identifying $\mathbb{R}^{n \times n} \cong \mathbb{R}^{n^{2}}$, $\mathbb{R}^{n^{2}\times n^{2}} \cong \mathbb{R}^{n^{4}}$, the map $A \mapsto A \otimes A : \mathbb{R}^{n \times n} \rightarrow  \mathbb{R}^{n^{2} \times n^{2}}$ is polynomial in its coordinates, hence smooth. Thus, since $\GL(n) \hookrightarrow \mathbb{R}^{n \times n}$ is an open subset, it follows that $\Phi : \GL(n) \rightarrow \mathbb{R}^{n^{2} \times n^{2}}$ is smooth by restriction of the domain \cite[Theorem 5.27]{Lee_diff_geom:book:2013}. But that $\Phi(\GL(n)) \subset \GL^{+}(n^{2})$ follows from \ref{prop:kron_propts_det}), so since $\GL^{+}(n^{2}) \hookrightarrow \GL(n^{2}) \hookrightarrow \mathbb{R}^{n^{2}\times n^{2}}$, we may then restrict the codomain as well \cite[Theorem 5.29]{Lee_diff_geom:book:2013}, yielding $\Phi : \GL(n) \rightarrow \GL^{+}(n^{2})$ is smooth. 
%
%Finally, in the case $n$ is odd, since $\left. \Phi \right|_{\GL^{+}(n)} : \GL^{+}(n) \rightarrow \GL^{+}(n^{2})$ is an injective Lie group homomorphism, its image $\left. \Phi \right|_{\GL^{+}(n)}(\GL^{+}(n))$ has a unique smooth structure such that $\left. \Phi \right|_{\GL^{+}(n)}(\GL^{+}(n)) \hookrightarrow \GL^{+}(n^{2})$ is a Lie subgroup and $\left. \Phi \right|_{\GL^{+}(n)} : \GL^{+}(n) \rightarrow \left. \Phi \right|_{\GL^{+}(n)}(\GL^{+}(n))$ is a Lie group isomorphism \cite[Proposition 7.17]{Lee_diff_geom:book:2013} 
%
The remaining claims are straightforward, noting that $-I \in \GL^{-}(n)$ if and only if $n$ is odd.
%
% ***********************
%
% QED
%
$\hfill\blacksquare$

% . Since $\GL^{+}(n) \hookrightarrow \GL(n)$ is an open subgroup, $\left. \Phi \right|_{\GL^{+}(n)} : \GL^{+}(n) \rightarrow \GL(n^{2})$ is an (injective) Lie group homomorphism. Consider then the Lie group homomorphism $\det \circ \left. \Phi \right|_{\GL^{+}(n)} : \GL^{+}(n) \rightarrow \mathbb{R} \backslash \{0\}$. Since $\det \circ \left. \Phi \right|_{\GL^{+}(n)}(I) = 1$, connectedness of $\GL^{+}(n)$ and continuity of $\det \circ \left. \Phi \right|_{\GL^{+}(n)}$ imply that $\det \circ \left. \Phi \right|_{\GL^{+}(n)}(\GL^{+}(n)) \subset (0, \infty)$ (indeed, equality holds). This says exactly that $\left. \Phi \right|_{\GL^{+}(n)}(\GL^{+}(n)) \subset \GL^{+}(n^{2})$.

% *************************************************************************
%
% PROPOSITION: SYMMETRIC KRONECKER PRODUCT PROPERTIES
%
% *************************************************************************

\begin{proposition}[Symmetric Kronecker Product Properties]\label{prop:skron_propts}
The symmetric Kronecker product has the following properties developed previously in the in the square-matrix case \cite{Todd_Toh_Tutuncu_skron:1998}, generalized here to rectangular matrices:

\begin{enumerate}[1S)]
	
	\item\label{prop:skron_propts_bilin} $\skron : \mathbb{R}^{m \times n} \times \mathbb{R}^{m \times n} \rightarrow \mathbb{R}^{\underline{m} \times \underline{n}}$ is bilinear.

	\item\label{prop:skron_propts_sym} $\skron$ is symmetric; i.e., $A \skron B = B \skron A$, $A, B \in \mathbb{R}^{m \times n}$.

	\item\label{prop:skron_propts_mixed_vec_prod} $(A \skron B) \svec(\pi(C)) = \svec(\pi(B \pi(C) A^{T}))$, $A, B \in \mathbb{R}^{m \times n}$, $C \in \mathbb{R}^{n \times n}$.
	
	\item\label{prop:skron_propts_transpose} $(A \skron B)^{T} = A^{T} \skron B^{T}$, $A, B \in \mathbb{R}^{m \times n}$.
	
	\item\label{prop:skron_propts_inv} $(A \skron A)^{-1} = A^{-1} \otimes A^{-1}$, $A \in \GL(n)$. However,  $(A \skron B)^{-1} \neq A^{-1} \skron B^{-1}$ for $A, B \in \GL(n)$, in general. Indeed, $A, B \in \GL(n)$ does not imply $A \skron B \in \GL(\underline{n})$.
	
	\item\label{prop:skron_propts_mixed_prod} 
	\begin{enumerate}[a)]
	
		\item $(A \skron B)(C \skron D) = \frac{1}{2} \left( AC \skron BD + AD \skron BC \right)$, $A, B \in \mathbb{R}^{m \times n}$, $C, D \in \mathbb{R}^{n \times p}$.

		\item $(A \skron B)(C \skron C) = AC \skron BC$, $A, B \in \mathbb{R}^{m \times n}$, $C \in \mathbb{R}^{n \times p}$.
		
		\item $(C \skron C)(A \skron B) = CA \skron CB$, $A, B \in \mathbb{R}^{m \times n}$, $C \in \mathbb{R}^{p \times m}$.
	 
	\end{enumerate}

	\item\label{prop:skron_propts_eig}
	\begin{enumerate}[a)]
	
		\item For a square matrix $A \in \mathbb{R}^{n \times n}$, if $\sigma(A) = \{\lambda_{i} \mid i = 1, \dots, n\}$, then $\sigma(A \skron A) = \{ \lambda_{i} \lambda_{j} \mid 1 \leq i \leq j \leq n\}$. Furthermore, if $x_{i}, x_{j} \in \mathbb{C}^{n}$ are eigenvectors corresponding to the eigenvalues $\lambda_{i}, \lambda_{j}$ of $A$, respectively, then $x_{i} \skron x_{j}$ is an eigenvector corresponding to the eigenvalue $\lambda_{i} \lambda_{j}$ of $A \skron A$.

		\item Suppose that $A, B \in \mathbb{R}^{n \times n}$ are simultaneously diagonalizable with common basis of eigenvectors $\{x_{i}\}_{i=1}^{n}$. If $\sigma(A) = \{\lambda_{i} \mid i = 1, \dots, n\}$ and $\sigma(B) = \{\mu_{j} \mid j = 1, \dots, n\}$ are the eigenvalues of $A$ and $B$ corresponding to the respective eigenvectors $\{x_{i}\}_{i=1}^{n}$, then $\sigma(A \skron B) = \left\{ \frac{1}{2} (\lambda_{i} \mu_{j} + \lambda_{j} \mu_{i} ) \mid 1 \leq i \leq j \leq n \right\}$. Furthermore, $x_{i} \skron x_{j}$ is an eigenvector corresponding to the eigenvalue $\frac{1}{2} (\lambda_{i} \mu_{j} + \lambda_{j} \mu_{i})$ of $A \skron B$.

		\item\label{prop:skron_propts_eig_shared_evecs} Suppose that $A, B \in \mathbb{R}^{n \times n}$ share two eigenvectors $x, y \in \mathbb{C}^{n}$. If $A x = \lambda_{1} x$, $B x = \mu_{1} x$, $A y = \lambda_{2} y$, $B y = \mu_{2} y$, then $x \skron y$ is an eigenvector of $A \skron B$ corresponding to the eigenvalue $\frac{1}{2} (\lambda_{1} \mu_{2} + \lambda_{2} \mu_{1})$.
	
	\end{enumerate}	

	\item\label{prop:skron_propts_ApI_sym} $A \skron I$ is symmetric if and only if $A$ is, $A \in \mathbb{R}^{n \times n}$.

	\item\label{prop:skron_propts_sym_pd} If $A, B \in \mathbb{S}^{n}$ are symmetric positive definite, then so is $A \skron B$.

    \item\label{prop:skron_propts_zero} $A \skron B = 0$ if and only if at least one $A, B = 0$, $A, B \in \mathbb{R}^{m \times n}$.

\end{enumerate}

\vspace*{0.125in}

\noindent In addition, the following newly-proved results are essential for the dEIRL MEE framework developed subsequently:

\vspace*{0.125in}

\begin{enumerate}[1S)]

    % Set counter
    \setcounter{enumi}{10}
 
	\item\label{prop:skron_propts_det} $\det(A \skron A) = \det(A)^{n+1}$, $A \in \mathbb{R}^{n \times n}$.

	\item\label{prop:skron_propts_diag} For $A, B \in \mathbb{R}^{n \times n}$, if $A, B$ are diagonal, then $A \skron B$ is diagonal. If $A, B$ are nonzero on each diagonal entry and $A \skron B$ is diagonal, then $A, B$ are diagonal.
	
	\item\label{prop:skron_propts_ApB_I} For $A, B \in \mathbb{R}^{n \times n}$, $A \skron B = I_{\underline{n}}$ if and only if $A = \lambda I_{n}$, $B = \frac{1}{\lambda} I_{n}$ for some $\lambda \neq 0$.
	
	\item\label{prop:skron_propts_LG_homo} The map $\underline{\Phi} : \GL(n) \rightarrow \GL(\underline{n})$, 
	\begin{align}
		\underline{\Phi}(A) 
		= 
		A \skron A,
		\qquad 
		A \in \GL(n),
		\label{eq:ul_Phi_def}
	\end{align}
 	is a Lie group homomorphism with $\ker \underline{\Phi} = \{\pm I\}$. 
	$\left. \underline{\Phi} \right|_{\GL^{+}(n)} : \GL^{+}(n) \rightarrow \GL^{+}(\underline{n})$ is an injective Lie group homomorphism if and only if $n$ is odd. 
	In the case $n$ is odd, $\underline{\Phi}(\GL(n)) = \underline{\Phi}(\GL^{+}(n)) \hookrightarrow \GL^{+}(\underline{n})$ is connected. In the case $n$ is even, $\underline{\Phi}(\GL(n))$ has two connected components $\underline{\Phi}(\GL^{+}(n)) \hookrightarrow \GL^{+}(\underline{n})$, $\underline{\Phi}(\GL^{-}(n)) \hookrightarrow \GL^{-}(\underline{n})$.

\end{enumerate}

\end{proposition}

% *************************************************************************
%
% PROOF
%
% *************************************************************************
%
\textit{Proof:} 
Aside from \ref{prop:skron_propts_eig}Sc), \ref{prop:skron_propts_bilin}S)--\ref{prop:skron_propts_sym_pd}S) were proved in \cite{Todd_Toh_Tutuncu_skron:1998} in the square-matrix case. Here, we generalize \ref{prop:skron_propts_bilin}S)--\ref{prop:skron_propts_transpose}S), \ref{prop:skron_propts_mixed_prod}S) to the rectangular-matrix case, and the arguments are similar. \ref{prop:skron_propts_mixed_vec_prod}S), in particular, follows from the commutative diagram (\ref{diag:svec}).

\ref{prop:skron_propts_eig}Sa), \ref{prop:skron_propts_eig}Sb) were originally proved in \cite{Todd_Toh_Tutuncu_skron:1998} and are well-understood, but because prior works on the symmetric Kronecker product define it only as an operation on square matrices, they have missed that $x_{i} \skron x_{j}$ constitute the eigenvectors of $A \skron B$ -- an important and intuitive result paralleling that of the usual Kronecker product (cf. Proposition \ref{prop:kron_propts} \ref{prop:kron_propts_eig})). \ref{prop:skron_propts_eig}Sb) was proved in \cite{Todd_Toh_Tutuncu_skron:1998} in the case of commuting square matrices $A, B \in \mathbb{S}^{n}$, but simultaneous diagonalizability is the key property enabling this result. Underpinning the arguments in \ref{prop:skron_propts_eig}Sa) and \ref{prop:skron_propts_eig}Sb) is \ref{prop:skron_propts_eig}Sc), which we prove here because it will be illustrative subsequently. With all terms as in the hypotheses of \ref{prop:skron_propts_eig}Sc), we first note the subtlety that $x, y \neq 0$ implies $x \skron y \neq 0$, by \ref{prop:skron_propts_zero}S) (proven below, independently of this result). Next, applying the \emph{now-generalized} mixed product identity \ref{prop:skron_propts_mixed_prod}S), we have
\begin{align}
	(A \skron B)(x \skron y)
	&=
	\frac{1}{2} \left( Ax \skron By + Ay \skron Bx \right)
	\nonumber
	\\
	&=
	\frac{1}{2} (\lambda_{1} \mu_{2} + \lambda_{2} \mu_{1}) \; x \skron y.
	\label{eq:skron_shared_evec}
\end{align}

The authors are unaware of \ref{prop:skron_propts_det}S)--\ref{prop:skron_propts_LG_homo}S) being proved previously. \ref{prop:skron_propts_det}S) follows from \ref{prop:skron_propts_eig}Sa).
For \ref{prop:skron_propts_zero}S), \ref{prop:skron_propts_diag}S)--\ref{prop:skron_propts_LG_homo}S), we employ the indexing maps $r$ (\ref{eq:rj_def}) and $c$ (\ref{eq:cj_def}), which together with the mixed product identity \ref{prop:skron_propts_mixed_prod}S) yield the symmetric Kronecker product indexing identity (\ref{eq:skron_index_id}). Straightforward application of (\ref{eq:skron_index_id}) yields \ref{prop:skron_propts_zero}S), \ref{prop:skron_propts_diag}S), and \ref{prop:skron_propts_ApB_I}S).
\begin{table*}
\begin{minipage}{1.0\textwidth}	
\centering	
\begin{align}
	(A \skron B)_{i,j}
	=
	\begin{cases}
		a_{r_{m}(i),r_{n}(j)} b_{r_{m}(i),r_{n}(j)}, & r_{m}(i) = c_{m}(i), \; r_{n}(j) = c_{n}(j),
		\\
		\frac{\sqrt{2}}{2} \left( a_{r_{m}(i),r_{n}(j)} b_{r_{m}(i),c_{n}(j)} + a_{r_{m}(i),c_{n}(j)} b_{r_{m}(i),r_{n}(j)} \right), & r_{m}(i) = c_{m}(i), \; r_{n}(j) < c_{n}(j),
		\\
		\frac{\sqrt{2}}{2} \left( a_{r_{m}(i),r_{n}(j)} b_{c_{m}(i),r_{n}(j)} + a_{c_{m}(i),r_{n}(j)} b_{r_{m}(i),r_{n}(j)} \right), & r_{m}(i) < c_{m}(i), \; r_{n}(j) = c_{n}(j),
		\\
		\frac{1}{2} \big( a_{r_{m}(i),r_{n}(j)} b_{c_{m}(i),c_{n}(j)} + a_{r_{m}(i),c_{n}(j)} b_{c_{m}(i),r_{n}(j)} 
		\\
		\qquad
		+ a_{c_{m}(i),r_{n}(j)} b_{r_{m}(i),c_{n}(j)} + a_{c_{m}(i),c_{n}(j)} b_{r_{m}(i),r_{n}(j)} \big), & r_{m}(i) < c_{m}(i), \; r_{n}(j) < c_{n}(j),		
	\end{cases}
	\qquad
	i = 1, \dots, \underline{m}, \; j = 1, \dots, \underline{n}.
	\label{eq:skron_index_id}
\end{align}
\end{minipage}	
\end{table*}
Finally, \ref{prop:skron_propts_LG_homo}S) follows from \ref{prop:skron_propts_diag}S) and \ref{prop:skron_propts_ApB_I}S) in an analogous argument to the one presented in the proof of Proposition \ref{prop:kron_propts} \ref{prop:kron_propts_LG_homo}). 
%
% ***********************
%
% QED
%
$\hfill\blacksquare$

% *************************************************************************
%
% REMARK: SKRON EIG
%
% *************************************************************************
%
\begin{remark}[On the Eigenstructure of the Symmetric Kronecker Product]\label{rk:skron_eig}
Equation (\ref{eq:skron_shared_evec}) elucidates a key issue surrounding the eigenstructure of the symmetric Kronecker product: In general, given eigenvectors $A x = \lambda_{1} x$, $B y = \mu_{2} y$ of $A, B \in \mathbb{R}^{n \times n}$, the first term in the expansion $Ax \skron By = \lambda_{1} \mu_{2} \; x \skron y$ always factors in the desired fashion. Yet, the second term $Ay \skron Bx = Bx \skron Ay$ need not be a scalar multiple of $x \skron y$, since $x$ is not an eigenvector of $B$ and $y$ is not an eigenvector of $A$, in general. Naturally, this makes the eigenstructure of the symmetric Kronecker product a significantly more complicated object of study than that of the usual Kronecker product, cf. \cite{Todd_Toh_Tutuncu_skron:1998,Kalantarova_PhD_thesis_skron:2019,Kalantarova_Tuncel_skron:2021}.

As a note, the eigenstructure results of Proposition \ref{prop:skron_propts_eig}S) require the symmetric Kronecker product as a map on complex matrices (specifically, when eigenvectors are complex-valued). As is the case with the standard Kronecker product, the necessary results may developed for the complex case. Following the practice of previous works \cite{Alizadeh_Haeberly_Overton_skron:1998,Todd_Toh_Tutuncu_skron:1998,Tuncel_Wolkowicz_skron:2005,Kalantarova_Tuncel_skron:2021,Kalantarova_PhD_thesis_skron:2019}, we avoid carrying out this process explicitly here to maintain scope.
\end{remark}

% *************************************************************************
%
% REMARK: A, B INVERTIBLE =/> skron(A,B) INVERTIBLE
%
% *************************************************************************
%
\begin{remark}\label{rk:AsB_not_inv}
For a counterexample illustrating the point of Proposition \ref{prop:skron_propts} \ref{prop:skron_propts_inv}S), consider $A = \diag(1, -1), B = I_{2} \in \GL(2)$. Then $A \skron B = \frac{1}{2} A \skrons A = \diag(1, 0, -1) \notin \GL(\underline{2})$. The key here is that $A$ possesses eigenvalues $\sigma(A) = \{\pm 1\}$ symmetric with respect to the origin (cf. Proposition \ref{prop:eig_skrons}). Note further on this point that $\sigma(A \skrons A) = \{ 1 + 1, 1 - 1, -1 -1 \}$.
\end{remark}

% *************************************************************************
%
% REMARK: DIAGONAL SKRON ID
%
% *************************************************************************
%
\begin{remark}
The strengthened hypotheses for the converse direction of Proposition \ref{prop:skron_propts} \ref{prop:skron_propts_diag}S) in relation to Proposition \ref{prop:kron_propts} \ref{prop:kron_propts_diag}) are necessary. Indeed, in the case $n = 2$, consider $A = e_{2} e_{1}^{T} \in \mathbb{R}^{2 \times 2}$. Then  $A, A^{T} \neq 0$, and neither of these matrices are diagonal, yet $A \skron A^{T} = \texttt{diag}(0, \frac{1}{2}, 0)$ is diagonal. Note that $A, A^{T}$ are zero on their diagonals.
\end{remark}

% *************************************************************************
%
% REMARK: LIE GROUP HOMO
%
% *************************************************************************
%
\begin{remark}[Lie Group Homomorphisms $\Phi$, $\underline{\Phi}$]
The Lie Group homomorphism $\underline{\Phi}$ in Proposition \ref{prop:skron_propts} \ref{prop:skron_propts_LG_homo}S) is relevant to the MEE framework developed in Section \ref{sec:prescaling}. To maintain subsequent emphasis on the symmetric Kronecker product algebra, we will after this section avoid labeling this map explicitly.
We have included construction of its Kronecker product counterpart $\Phi$ in Proposition \ref{prop:kron_propts} \ref{prop:kron_propts_LG_homo}) for completeness. By virtue of the bilinearity of the (symmetric) Kronecker product, these homomorphisms are homogeneous of degree two.
For intuition, consider the case $n = 1$. Then $\underline{n} = 1$, $r_{1} \equiv 1$ (\ref{eq:rj_def}), $c_{1} \equiv 1$ (\ref{eq:cj_def}), $\{E_{i}\}_{i=1}^{\underline{1}} = \{1\}$ (\ref{eq:Sn_ONB}), and $W_{1} = 1$ (\ref{eq:W_def}). In all, $\otimes = \skron$ are both given by scalar multiplication, and $\Phi(a) = \underline{\Phi}(a) = a^{2}$ (we will thus focus on $\underline{\Phi}$). Here, $\underline{\Phi}: \GL(1) = \mathbb{R} \backslash \{0\} \rightarrow \GL^{+}(1) = (0, \infty)$. This is a group homomorphism: $\underline{\Phi}(ab) = abab = aa bb = \underline{\Phi}(a) \underline{\Phi}(b)$, which is polynomial in the global coordinate on $\mathbb{R} \backslash \{0\} \hookrightarrow \mathbb{R}$ and on $(0, \infty) \hookrightarrow \mathbb{R}$, hence smooth. Note also that $\underline{\Phi}(a) = a^{2} = 1$ if and only if $a \in \{\pm 1\}$. Finally, $\left. \underline{\Phi} \right|_{\GL^{+}(1)}: \GL^{+}(1) = (0, \infty) \rightarrow \GL^{+}(\underline{1}) = (0, \infty)$ is a Lie group isomorphism onto its image $\left. \underline{\Phi} \right|_{\GL^{+}(1)}((0, \infty)) = \underline{\Phi}((0, \infty)) = (0, \infty)$ (a connected subgroup of $\GL^{+}(\underline{1}) = (0,\infty)$); in particular, the map $a \mapsto a^{2} : (0, \infty) \rightarrow (0, \infty)$ is a diffeomorphism. 

In the above, $\left. \underline{\Phi} \right|_{\GL^{+}(1)}: \GL^{+}(1) \rightarrow \GL^{+}(\underline{1})$ is a Lie group isomorphism in its own right. However, in the case $n > 1$, $\left. \underline{\Phi} \right|_{\GL^{+}(n)}: \GL^{+}(n) \rightarrow \GL^{+}(\underline{n})$ is not a Lie group isomorphism. In the case $n$ is even, it fails to be injective. Meanwhile, for all $n > 1$, $\underline{\Phi}$ fails to be onto $\GL^{+}(\underline{n})$. For otherwise $\underline{\Phi}$ would be a surjective map of constant rank, hence a submersion by the global rank theorem \cite[Theorem 4.14]{Lee_diff_geom:book:2013}; i.e., $\rank(\underline{\Phi}) = \underline{n}$ -- a contradiction of the fact that $\rank(\underline{\Phi}) \leq \min \{n, \underline{n}\} = n < \underline{n}$ always. A similar argument prevails for $\Phi$.
\end{remark}

Having discussed the (symmetric) Kronecker product, we now move on to the (symmetric) Kronecker sum. We first recall the spectral result in the standard case:

% *************************************************************************
%
% PROPOSITION: KRONECKER SUM EIGENSTRUCTURE
%
% *************************************************************************
%
\begin{proposition}[Eigenstructure of The Kronecker Sum \text{\cite[Theorem 4.4.5]{Horn_matrix_analysis:book:1991}}]\label{prop:kron_sum_eig}
For square matrices $A \in \mathbb{R}^{n \times n}$ and $B \in \mathbb{R}^{m \times m}$, if $\sigma(A) = \{\lambda_{i} \mid i = 1, \dots, n\}$ and $\sigma(B) = \{\mu_{j} \mid j = 1, \dots, m\}$, then $\sigma(A \oplus B) = \{ \lambda_{i} + \mu_{j} \mid i = 1, \dots, n, \, j = 1, \dots, m\}$. Furthermore, if $x_{i} \in \mathbb{C}^{n}$, $y_{j} \in \mathbb{C}^{m}$ are eigenvectors corresponding to the eigenvalues $\lambda_{i}$ of $A$ and $\mu_{j}$ of $B$, respectively, then $x_{i} \otimes y_{j}$ is an eigenvector corresponding to the eigenvalue $\lambda_{i} + \mu_{j}$ of $A \oplus B$.
\end{proposition}

While the eigenstructure of the Kronecker sum is quite intuitive, the eigenstructure of the symmetric Kronecker sum is more complicated, owing to the complications inherited from the symmetric Kronecker product (cf. Remark \ref{rk:skron_eig}). In the simultaneously-diagonalizable case, the result of Proposition \ref{prop:skron_propts_eig}Sb), developed originally in \cite{Todd_Toh_Tutuncu_skron:1998}, may be applied to the symmetric Kronecker sum as follows:

% *************************************************************************
%
% PROPOSITION: SYMMETRIC KRONECKER SUM EIGENSTRUCTURE
%
% *************************************************************************

\begin{proposition}[Eigenstructure of The Symmetric Kronecker Sum (Simultaneously Diagonalizable Case)]\label{prop:eig_skrons_simult_diag}
Suppose that $A, B \in \mathbb{R}^{n \times n}$ are simultaneously diagonalizable with common basis of eigenvectors $\{x_{i}\}_{i=1}^{n}$. If $\sigma(A) = \{\lambda_{i} \mid i = 1, \dots, n\}$ and $\sigma(B) = \{\mu_{j} \mid j = 1, \dots, n\}$ are the eigenvalues of $A$ and $B$ corresponding to the respective eigenvectors $\{x_{i}\}_{i=1}^{n}$, then $\sigma(A \skrons B) = \left\{ \frac{1}{2} (\lambda_{i} + \mu_{i} + \lambda_{j} + \mu_{j} ) \mid 1 \leq i \leq j \leq n \right\}$. Furthermore, $x_{i} \skron x_{j}$ is an eigenvector corresponding to the eigenvalue $\frac{1}{2} (\lambda_{i} + \mu_{i} + \lambda_{j} + \mu_{j})$ of $A \skrons B$.
\end{proposition}

For our purposes, Proposition \ref{prop:eig_skrons_simult_diag} is too restrictive. The following property will be useful shortly:

% *************************************************************************
%
% PROPOSITION: SYMMETRIC KRONECKER SUM EIGENSTRUCTURE
%
% *************************************************************************

\begin{lemma}[Partial Eigenstructure of The Symmetric Kronecker Sum]\label{lem:eig_skrons_partial}
Suppose that $A, B \in \mathbb{R}^{n \times n}$ share two eigenvectors $x, y \in \mathbb{C}^{n}$. If $A x = \lambda_{1} x$, $B x = \mu_{1} x$, $A y = \lambda_{2} y$, $B y = \mu_{2} y$, then $x \skron y$ is an eigenvector of $A \skrons B$ corresponding to the eigenvalue $\frac{1}{2} (\lambda_{1} + \mu_{1} + \lambda_{2} + \mu_{2})$.
\end{lemma}

% ***********************
%
% PROOF
%
\textit{Proof:} 
Follows from Proposition \ref{prop:skron_propts} \ref{prop:skron_propts_eig}Sc).
$\hfill\blacksquare$

Lemma \ref{lem:eig_skrons_partial} allows us to enumerate the eigenstructure of $A \skrons A$, a special case relevant to ALEs.

% *************************************************************************
%
% PROPOSITION: SYMMETRIC KRONECKER SUM EIGENSTRUCTURE
%
% *************************************************************************

\begin{proposition}[Eigenstructure of The Symmetric Kronecker Sum $A \skrons A$]\label{prop:eig_skrons}
For a square matrix $A \in \mathbb{R}^{n \times n}$, if $\sigma(A) = \{\lambda_{i} \mid i = 1, \dots, n\}$, then $\sigma(A \skrons A) = \{ \lambda_{i} + \lambda_{j} \mid 1 \leq i \leq j \leq n\}$. Furthermore, if $x_{i}, x_{j} \in \mathbb{C}^{n}$ are eigenvectors corresponding to the eigenvalues $\lambda_{i}, \lambda_{j}$ of $A$, respectively, then $x_{i} \skron x_{j}$ is an eigenvector corresponding to the eigenvalue $\lambda_{i} + \lambda_{j}$ of $A \skrons A$.
\end{proposition}

% *************************************************************************
%
% PROOF
%
% *************************************************************************
%
\textit{Proof:} 
Follows from Lemma \ref{lem:eig_skrons_partial}.
$\hfill\blacksquare$

Having discussed eigenstructure, we move on to the key exponentiation identity involving the (symmetric) Kronecker sum:

% *************************************************************************
%
% PROPOSITION: SYMMETRIC KRONECKER SUM EIGENSTRUCTURE
%
% *************************************************************************

\begin{proposition}[Exponentiation of the Kronecker Sum {\cite{Horn_matrix_analysis:book:1991}}]\label{prop:exp_krons}
Let $A \in \mathbb{R}^{m \times m}$, $B \in \mathbb{R}^{n \times n}$ be given.
\begin{enumerate}[1)]

    \item $(A \otimes I)^{k} = A^{k} \otimes I$, and $(I \otimes B)^{k} = I \otimes B^{k}$, $k \geq 0$.

    \item $\exp(A \oplus B) = \exp(A) \otimes \exp(B)$.
    
\end{enumerate}
\end{proposition}

The analogue holds for the symmetric Kronecker sum in the case $A = B$:

% *************************************************************************
%
% PROPOSITION: SYMMETRIC KRONECKER SUM EIGENSTRUCTURE
%
% *************************************************************************

\begin{proposition}[Exponentiation of the Symmetric Kronecker Sum]\label{prop:exp_skrons}
Let $A, B \in \mathbb{R}^{n \times n}$ be given.
\begin{enumerate}[1S)]

    \item\label{prop:exp_skrons_AsIk} $(A \skron I)^{k} = (I \skron A)^{k}$ is given by the following binomial expansion
    \begin{align}
        (A \skron I)^{k} 
        =
        \frac{1}{2^{k}}
        \sum\limits_{i=0}^{k} {k \choose i} A^{k-i} \skron A^{i},  
        \qquad k \geq 0.
        \label{eq:skron_AsIk}
    \end{align}
    \item\label{prop:exp_skrons_exp_id} $\exp(A \skrons A) = \exp(A) \skron \exp(A)$. However, in general $\exp(A \skrons B) \neq \exp(A) \skron \exp(B)$.
    
\end{enumerate}
\end{proposition}

% *************************************************************************
%
% PROOF
%
% *************************************************************************
%
\textit{Proof:} 
Proving that (\ref{eq:skron_AsIk}) holds is a quick algebraic check following from the mixed product identity of Proposition \ref{prop:skron_propts} \ref{prop:skron_propts_mixed_prod}S). \ref{prop:exp_skrons_exp_id}S) follows from (\ref{eq:skron_AsIk}) after examining the partial sums of $\exp(A \skrons A)$ and $\exp(A) \skron \exp(A)$.
$\hfill\blacksquare$

% *************************************************************************
%
% REMARK: exp(skrons(A,B)) ~= skron(exp(A),exp(B)) 
%
% *************************************************************************
%
\begin{remark}
For a counterexample illustrating the point of Proposition \ref{prop:exp_skrons} \ref{prop:exp_skrons_exp_id}S), consider the same matrices as in Remark \ref{rk:AsB_not_inv}: $A = \diag(1, -1)$, $B = I_{2}$. Then 
\begin{align}
    \exp(A \skrons B) 
    &= 
    \diag(e^{2}, e, 1), 
    \nonumber
    \\
    \exp(A) \skron \exp(B) 
    &=
    \diag \left(e^{2}, \, \frac{e^{2} + 1}{2}, \, 1 \right).
\end{align}
\end{remark}

% ************************************************************************
% ************************************************************************
% ************************************************************************
%
% SUBSECTION: ALEs
%
% ************************************************************************
% ************************************************************************
% ************************************************************************

\subsection{Symmetric Kronecker Products in Algebraic Lyapunov Equations (ALEs)}\label{ssec:skron_ALEs}

As is well-known, the Kronecker product plays an important role in characterizing existence and uniqueness of solutions to ALEs \cite{Horn_matrix_analysis:book:1991}. We illustrate in this section that the symmetric Kronecker product algebra developed above also provides this same characterization under symmetric conditions. Substantively, the algebra is structurally identical to the standard case.

% *************************************************************************
%
% DEFINITION: ALGEBRAIC LYAPUNOV EQUATION
%
% *************************************************************************

\begin{definition}[Algebraic Lyapunov Equation (ALE)]
Given $A \in \mathbb{R}^{n \times n}, B \in \mathbb{R}^{n \times m}$, consider the following algebraic Lyapunov equation (ALE)
\begin{align}
	A^{T} X + X A + B
	=
	0.
	\label{eq:ALE_gen}
\end{align}
\end{definition}

% *************************************************************************
%
% PROPOSITION: ALGEBRAIC LYAPUNOV EQUATION SOLUTIONS
%
% *************************************************************************

\begin{proposition}[ALE Existence and Uniqueness of Solutions]\label{prop:ALE_existence_uniqueness}
Let $\sigma(A) = \{\lambda_{i} \mid i = 1, \dots, n\}$. There exists a unique solution $X \in \mathbb{R}^{n \times m}$ of the ALE (\ref{eq:ALE_gen}) if and only if $\lambda_{i} + \lambda_{j} \neq 0$ for all $1 \leq i, j \leq n$.
\end{proposition}

% ***********************
%
% PROOF
%
\textit{Proof:} 
This proof is quite standard; see, e.g., \cite{Horn_matrix_analysis:book:1991,Brewer_Kronecker_products:1978}. However, we include it here to illustrate structural parallels to the analogous results developed shortly for the symmetric Kronecker product.
Applying the identities in Proposition \ref{prop:kron_propts}, we see that (\ref{eq:ALE_gen}) is equivalent to
\begin{align}
	\vecc(A^{T} X + X A)
	=
	(A \oplus A)^{T} \vecc(X)
	=
	-\vecc(B).
	\label{eq:ALE_vec}
\end{align}
Thus, the ALE (\ref{eq:ALE_gen}) has a unique solution if and only if $(A \oplus A)^{T} \in \GL(n^{2})$. Applying Proposition \ref{prop:kron_sum_eig}, $\sigma((A \oplus A)^{T}) = \{\lambda_{i} + \lambda_{j} \mid i,j = 1, \dots, n\}$, from which the result follows.
$\hfill\blacksquare$

% *************************************************************************
%
% PROPOSITION: ALGEBRAIC LYAPUNOV EQUATION SOLUTIONS -- STABLE CASE
%
% *************************************************************************

\begin{proposition}[ALE Existence and Uniqueness of Solutions: Stable Systems \text{\cite[Proposition 5.2.1]{AAR_multivariable:book}}]\label{prop:ALE_existence_uniqueness_stable}
Suppose $A \in \mathbb{R}^{n \times n}$ is Hurwitz, and $Q \in \mathbb{S}^{n}$. Consider the ALE
\begin{align}
	A^{T} P + P A + Q
	=
	0.
	\label{eq:ALE}
\end{align}

\begin{enumerate}[1)]

	\item The unique solution is the symmetric matrix
	\begin{align}
		P
		=
		\int_{0}^{\infty} e^{A^{T} t} Q e^{A t} dt.
	\end{align}
	
	\item If $Q$ is positive (semi)definite, then $P$ is positive (semi)definite.
	
	\item If $Q$ is positive semidefinite, then $P$ is positive definite if and only if $(Q^{1/2}, A)$ is detectable.

\end{enumerate}

\end{proposition}

% *************************************************************************
%
% PROPOSITION: ALGEBRAIC LYAPUNOV EQUATION SOLUTIONS -- STABLE CASE
%
% *************************************************************************

\begin{remark}[Symmetric Kronecker Algebra of the ALE (\ref{eq:ALE})]
Consider the ALE (\ref{eq:ALE}). Applying Proposition \ref{prop:ALE_existence_uniqueness_stable}, we know $P \in \mathbb{S}^{n}$. We may then apply the symmetric Kronecker product algebra in Proposition \ref{prop:skron_propts}, yielding
\begin{align}
	\svec(A^{T} P + P A)
	=
	- \svec(Q).
	\label{eq:ALE_vec_skron}
\end{align}
Now, applying Proposition \ref{prop:skron_propts} \ref{prop:skron_propts_mixed_vec_prod}S), the left-hand-side of (\ref{eq:ALE_vec_skron}) becomes,
\begin{align}
	2 \svec(\pi(P A))
	=
	2 (A^{T} \skron I) \svec(P)
	=
	(A \skrons A)^{T} \svec(P).
\end{align}
Altogether, the ALE (\ref{eq:ALE}) is equivalent to the following:
\begin{align}
	\svec(A^{T} P + P A)
	=
	(A \skrons A)^{T} \svec(P)
	=
	- \svec(Q).
	\label{eq:ALE_svec}
\end{align}
The reader is encouraged to compare Equations (\ref{eq:ALE_vec}) and (\ref{eq:ALE_svec}), which precisely motivates our definition of the symmetric Kronecker sum $\skrons$ as the natural analogue to the Kronecker sum $\oplus$. The structural parallels extend further: Note by Proposition \ref{prop:eig_skrons} that $\sigma((A \skrons A)^{T}) = \{\lambda_{i} + \lambda_{j} \mid 1 \leq i \leq j \leq n\}$. Thus, in the case $Q \in \mathbb{S}^{n}$, the symmetric Kronecker sum may be used to characterize existence and uniqueness of solutions to the ALE (\ref{eq:ALE}) in an entirely similar argument to the one used in the proof of Proposition \ref{prop:ALE_existence_uniqueness}. Here, the square-symmetric nature of the matrix $Q \in \mathbb{S}^{n}$ has enabled an effective reduction is dimensionality of the problem from $n^{2}$ to $\underline{n}$.

\end{remark}

%\FloatBarrier

%\fi				% Comment this line and \iffalse to include section

% *************************************************************************
%
% PRESCALING
%
% *************************************************************************

%\iffalse			% Comment this line and \fi to include section

% ************************************************************************
% ************************************************************************
% ************************************************************************
%
% PRESCALING
%
% ************************************************************************
% ************************************************************************
% ************************************************************************

\section{Modulation-Enhanced Excitation (MEE) Framework}\label{sec:prescaling}

% *************************************************************************
%
% KLEINMAN'S
%
% *************************************************************************

Let a decentralized loop $1 \leq j \leq N$ be given, and suppose that $K_{0,j} \in \mathbb{R}^{m_{j} \times n_{j}}$ is such that $A_{jj} - B_{jj} K_{0,j}$ is Hurwitz in loop $j$. 
We may then apply Kleinman's algorithm (Section \ref{sec:problem_formulation}), yielding sequences $\{P_{i,j}\}_{i=0}^{\infty}$ in $\mathbb{R}^{n_{j} \times n_{j}}$ and $\{K_{i,j}\}_{i=0}^{\infty}$ in $\mathbb{R}^{m_{j} \times n_{j}}$ from the ALE 
\begin{align}
	A_{i,j}^{T} P_{i,j} + P_{i,j} A_{i,j} + Q_{i,j}
	=
	0.
	\label{eq:Kleinman_LE_decoupled}
\end{align}
where the matrices $A_{i,j}$ and $Q_{i,j}$ are given by (\ref{eq:nonlin_sys_dirl_rewritten}) and (\ref{eq:DIRL_nonlin_lsq_b}), respectively. We have seen, vis. (\ref{eq:ALE_svec}), that the ALE (\ref{eq:Kleinman_LE_decoupled}) is equivalent to the following vectorized ALE regression
\begin{align}
	(A_{i,j} \skrons A_{i,j})^{T} \svec(P_{i,j})
	=
	- \svec(Q_{i,j}).
	\label{eq:ALE_svec_Kleinman}
\end{align}
\noindent Now, suppose $S = \texttt{diag}(S_{1}, \dots, S_{N}) \in \textnormal{GL}(n)$, $S_{j} \in \GL(n_{j})$ $(j = 1, \dots, N)$, is any nonsingular coordinate transformation $\tilde{x} = S x$, partitioned in the decentralized form
\begin{align}
	\tilde{x}_{j} 
	=
	S_{j} x_{j}.
\end{align}
This induces the following transformed LQR problem in loop $j$, associated with the quadruple $(\tilde{A}_{jj}, \tilde{B}_{jj}, \tilde{Q}_{j}, R_{j})$, where
\begin{align}
	\tilde{A}_{jj} = S_{j} A_{jj} S_{j}^{-1},
	\;\;
	\tilde{B}_{jj} = S_{j} B_{jj},
	\;\;
	\tilde{Q}_{j} = S_{j}^{-T} Q_{j} S_{j}^{-1}.
	\label{eq:prescaled_terms_Kleinman}
\end{align}
\noindent By similarity, the controller $\tilde{K}_{i,j} = K_{i,j} S_{j}^{-1}$ is such that $\tilde{A}_{i,j} = \tilde{A}_{jj} - \tilde{B}_{jj} \tilde{K}_{i,j}$ is Hurwitz. This motivates the following modulated ALE
\begin{align}
	\tilde{A}_{i,j}^{T} \tilde{P}_{i,j} + \tilde{P}_{i,j} \tilde{A}_{i,j} + \tilde{Q}_{i,j}
	=
	0.
	\label{eq:Kleinman_LE_decoupled_transformed}
\end{align}

Modulation by nonsingular coordinate transformations is common practice in the study of matrix equations, oftentimes offering significant theoretical/numerical advantages for purposes of solving \cite{Horn_matrix_analysis:book:1991}. At this point, two questions are natural: 1) How do the original sequences $\{P_{i,j}\}_{i=0}^{\infty}$, $\{K_{i,j}\}_{i=0}^{\infty}$ output by Kleinman's algorithm relate to the modulated sequences $\{\tilde{P}_{i,j}\}_{i=0}^{\infty}$, $\{\tilde{K}_{i,j}\}_{i=0}^{\infty}$? Noting by Theorem \ref{thm:DIRL_nonlin_Kleinman_equivalence} that dEIRL and Kleinman's algorithm are equivalent, this first question also addresses the relations between the respective sequences produced by dEIRL. And, 2) How does prescaling interact with the symmetric Kronecker product algebra developed in Section \ref{sec:skron}? That is, how does prescaling affect the terms in the ALE regression (\ref{eq:ALE_svec_Kleinman}) and the dEIRL regression (\ref{eq:DIRL_nonlin_lsq}), and what structural parallels exist between the two?

% ************************************************************************
% ************************************************************************
% ************************************************************************
%
% SUBSECTION: KLEINMAN'S AND PRESCALING
%
% ************************************************************************
% ************************************************************************
% ************************************************************************

\subsection{Kleinman's Algorithm \& Modulation}\label{ssec:prescaling_Kleinman}

% *************************************************************************
%
% TABLE: PRESCALING PROPERTIES
%
% *************************************************************************

\begin{table*}[ht]
	\caption{Kleinman's Algorithm \& dEIRL: Symmetric Kronecker Product Algebraic Structure under Modulation}
	\begin{minipage}{1.0\textwidth}	
	\centering
	\begin{tabular}{|c||c|c||c|c|}
	        \hline
	        \multirow{2}{*}{Term} & \multicolumn{2}{c||}{Kleinman Loop $j$}  & \multicolumn{2}{c|}{dEIRL Loop $j$}
	        \\
	        \hhline{~||-|-||-|-|}
	        & Original & w/ Modulation  &  Original & w/ Modulation (MEE)   
	        \\
	        \hline
	        % ***** EIRL *****
	        \hline\Tstrut    
	        Dynamics & $A_{jj}$, $B_{jj}$ (\ref{eq:sys_lin_2x2})  & $S_{j} A_{jj} S_{j}^{-1}$, $S_{j} B_{jj}$ (\ref{eq:prescaled_terms_Kleinman}) & $f_{j}$, $g_{j}$ (\ref{eq:sys_nonlin_2x2}) & $S_{j} \circ f_{j} \circ S^{-1}$, $S_{j} \circ g_{j} \circ S^{-1}$ (\ref{eq:prescaled_terms_dEIRL})
	        \\
	        \hline\Tstrut
	        Cost Structure & $Q_{j}$, $R_{j}$ (\ref{eq:LQR_penalties_2x2}) & $S_{j}^{-T} Q_{j} S_{j}^{-1}$, $R_{j}$ (\ref{eq:prescaled_terms_Kleinman}) & $Q_{j}$, $R_{j}$ (\ref{eq:LQR_penalties_2x2}) & $S_{j}^{-T} Q_{j} S_{j}^{-1}$, $R_{j}$ (\ref{eq:prescaled_terms_Kleinman})
	        \\
	        \hline\Tstrut	       
	        ALE Solution & $P_{i,j}$ (\ref{eq:Kleinman_LE_decoupled}) & $S_{j}^{-T} P_{i,j} S_{j}^{-1}$ (\ref{eq:Pij_Kij_vs_tPij_tKij})  & $P_{i,j}$ (\ref{eq:Kleinman_LE_decoupled}) & $S_{j}^{-T} P_{i,j} S_{j}^{-1}$ (\ref{eq:Pij_Kij_vs_tPij_tKij})
	        \\
	        \hline\Tstrut	 
	        Controller & $K_{i,j}$ (\ref{eq:DIRL_controller_update}) & $K_{i,j} S_{j}^{-1}$ (\ref{eq:Pij_Kij_vs_tPij_tKij}) & $K_{i,j}$ (\ref{eq:DIRL_controller_update}) & $K_{i,j} S_{j}^{-1}$ (\ref{eq:Pij_Kij_vs_tPij_tKij})
	        \\
	        \hline	 	     
	        Regression  & \multirow{2}{*}{$(A_{i,j} \skrons A_{i,j})^{T}$ (\ref{eq:ALE_svec_Kleinman})}  & \multirow{2}{*}{$(A_{i,j} \skrons A_{i,j})^{T} (S_{j} \skron S_{j})^{T}$ (\ref{eq:ALE_svec_Kleinman_prescaled_equivalent})}  & \multirow{2}{*}{$\mathbf{A}_{i,j}$ (\ref{eq:DIRL_nonlin_lsq_A})} & \multirow{2}{*}{$\mathbf{A}_{i,j} (S_{j} \skron S_{j})^{T}$ (\ref{eq:DIRL_nonlin_lsq_Ab_orig_vs_prescaled})}
	        \\
	        \hhline{~||~|~||~|~|}
	        Matrix & & & &
	        \\
	        \hline	       
	        Regression & \multirow{2}{*}{$- \svec(Q_{i,j})$ (\ref{eq:ALE_svec_Kleinman})}  & \multirow{2}{*}{$- \svec(Q_{i,j})$ (\ref{eq:ALE_svec_Kleinman_prescaled_equivalent})}  & \multirow{2}{*}{$- I_{x_{j}, x_{j}} \svec(Q_{i,j})$ (\ref{eq:DIRL_nonlin_lsq_b})} & \multirow{2}{*}{$ - I_{x_{j}, x_{j}} \svec(Q_{i,j})$ (\ref{eq:DIRL_nonlin_lsq_Ab_orig_vs_prescaled})}
	        \\
	        \hhline{~||~|~||~|~|}
	        Target Vector & & & &	        
	        \\
	        \hline	 	           	            
	\end{tabular}
	\vspace{-0.3cm}	
	\end{minipage}	
	\label{tb:prescaling_propts}
\end{table*}

% *************************************************************************
%
% PROPOSITION: SOLUTION OF KLEINMAN'S (PRESCALED)
%
% *************************************************************************

\begin{theorem}[Kleinman's Algorithm: Modulation Invariance]\label{thm:Kleinman_prescaling_invariance}
$P_{i,j} \in \mathbb{S}^{n_{j}}$, $P_{i,j} > 0$ satisfies the ALE (\ref{eq:Kleinman_LE_decoupled}) if and only if $\tilde{P}_{i,j} = S_{j}^{-T} P_{i,j} S_{j}^{-1}$ satisfies the modulated ALE (\ref{eq:Kleinman_LE_decoupled_transformed}).
	
\end{theorem}

% *************************************************************************
%
% PROOF
%
% *************************************************************************
%
\textit{Proof:} 
We have seen, vis. (\ref{eq:ALE_svec}), that the modulated ALE (\ref{eq:Kleinman_LE_decoupled_transformed}) is equivalent to
\begin{align}
	(\tilde{A}_{i,j} \skrons \tilde{A}_{i,j})^{T} \svec(\tilde{P}_{i,j})
	=
	- \svec(\tilde{Q}_{i,j}).
	\label{eq:ALE_svec_Kleinman_prescaled}
\end{align}
Applying the symmetric Kronecker product algebra of Proposition \ref{prop:skron_propts}, we may expand (\ref{eq:ALE_svec_Kleinman_prescaled}) as
\begin{align}
	&
	(S_{j} \skron S_{j})^{-T} (A_{i,j} \skrons A_{i,j})^{T} (S_{j} \skron S_{j})^{T} \svec(\tilde{P}_{i,j})
	\nonumber
	\\
	&\qquad\qquad=
	- (S_{j} \skron S_{j})^{-T} \svec(Q_{i,j}).
\end{align}
By Proposition \ref{prop:skron_propts} \ref{prop:skron_propts_inv}S), we may multiply both sides by $(S_{j} \skron S_{j})^{T} \in \GL(\underline{n}_{j})$, yielding the equivalent regression
\begin{align}
	(A_{i,j} \skrons A_{i,j})^{T} (S_{j} \skron S_{j})^{T} \svec(\tilde{P}_{i,j})
	=
	- \svec(Q_{i,j}).
	\label{eq:ALE_svec_Kleinman_prescaled_equivalent}
\end{align}
However, from comparison of (\ref{eq:ALE_svec_Kleinman_prescaled_equivalent}) and the symmetric vectorization of the original ALE (\ref{eq:ALE_svec_Kleinman}), we conclude that $(S_{j} \skron S_{j})^{T} \svec(\tilde{P}_{i,j}) = \svec(P_{i,j})$. Applying Proposition \ref{prop:skron_propts} again,
\begin{align}
	(S_{j} \skron S_{j})^{T} \svec(\tilde{P}_{i,j})
	&=
	\svec(\pi(S_{j}^{T} \tilde{P}_{i,j} S_{j}))
	\nonumber
	\\
	&=
	\svec(S_{j}^{T} \tilde{P}_{i,j} S_{j}).
\end{align}
In all, $S_{j}^{T} \tilde{P}_{i,j} S_{j} = P_{i,j}$, implying the desired result. The reverse direction follows by a symmetric argument.
%
% ***********************
%
% QED
%
$\hfill\blacksquare$

We now have a powerful answer to question 1) posed above: Kleinman's algorithm (and hence the dEIRL algorithm) is invariant with respect to nonsingular state modulation in the sense that if the sequences $\{\tilde{P}_{i,j}\}_{i=0}^{\infty}$, $\{\tilde{K}_{i,j}\}_{i=0}^{\infty}$ are generated under the modulated problem with potentially-improved numerics, then the solution sequences $\{P_{i,j}\}_{i=0}^{\infty}$, $\{K_{i,j}\}_{i=0}^{\infty}$ of the original problem may be backed out by
\begin{align}
	P_{i,j} 
	= 
	S_{j}^{T} \tilde{P}_{i,j}  S_{j},
	\qquad
	K_{i,j} 
	=
	\tilde{K}_{i,j} S_{j}.
	\label{eq:Pij_Kij_vs_tPij_tKij}
\end{align}
Furthermore, the above proof also answers question 2) in the case of Kleinman's algorithm: The modulated ALE regression (\ref{eq:ALE_svec_Kleinman_prescaled}) is equivalent to (\ref{eq:ALE_svec_Kleinman_prescaled_equivalent}), in which we observe the that the original ALE regression matrix $(A_{i,j} \skrons A_{i,j})^{T} \in \GL(\underline{n}_{j})$ (\ref{eq:ALE_svec_Kleinman}) is multiplied on the right by the modulation matrix $(S_{j} \skron S_{j})^{T} \in \GL(\underline{n}_{j})$. The regression target vector $- \svec(Q_{i,j}) \in \mathbb{R}^{\underline{n}_{j}}$ is unchanged between the original regression (\ref{eq:ALE_svec_Kleinman}) and equivalent modulated regression (\ref{eq:ALE_svec_Kleinman_prescaled_equivalent}).

% ************************************************************************
% ************************************************************************
% ************************************************************************
%
% SUBSECTION: dEIRL AND PRESCALING
%
% ************************************************************************
% ************************************************************************
% ************************************************************************

\subsection{dEIRL \& Modulation: MEE Framework}\label{ssec:prescaling_dEIRL}

Now, consider the analogue in the dEIRL algorithm. Associate with the nonsingular coordinate transformation $S_{j} \in \textnormal{GL}(n_{j})$ the transformed problem $(\tilde{f}_{j}, \tilde{g}_{j}, \tilde{Q}_{j}, R_{j})$ in loop $j$, where
\begin{align}
	\tilde{f}_{j} = S_{j} \circ f_{j} \circ S^{-1},
	\quad
	\tilde{g}_{j} = S_{j} \circ g_{j} \circ S^{-1}.
	\label{eq:prescaled_terms_dEIRL}
\end{align}
This induces the following modulated dEIRL least-squares regression, analogous to (\ref{eq:DIRL_nonlin_lsq}), which we term the MEE regression for brevity:
\begin{align}
	\tilde{\mathbf{A}}_{i,j} \svec (\tilde{P}_{i,j})
	=
	\tilde{\mathbf{b}}_{i,j}.
	\label{eq:DIRL_nonlin_lsq_prescaled}
\end{align}

The symmetric Kronecker product algebraic results developed in Section \ref{sec:skron} are essential to the derivation of the MEE regression (\ref{eq:DIRL_nonlin_lsq_prescaled}). In particular:

% *************************************************************************
%
% PROPOSITION: RELEVANT OPERATORS
%
% *************************************************************************

\begin{proposition}\label{prop:dEIRL_ops_propts}
The operations $\delta_{x,y}$ (\ref{eq:delta_xy_def}) and $I_{x, y}$ (\ref{eq:I_xy_def}) satisfy the following:
\begin{enumerate}[1)]

	\item $\delta_{Ax,Ay} = \delta_{x,y} (A \skron A)^{T}$, $A \in \mathbb{R}^{m \times n}$.
	
	\item $I_{Ax,Ay} = I_{x,y} (A \skron A)^{T}$, $A  \in \mathbb{R}^{m \times n}$.
	
	\item $I_{Ax,Bx} = I_{x,x} (A \skron B)^{T}$, $A, B \in \mathbb{R}^{m \times n}$.

\end{enumerate}

\end{proposition}

% ***********************
%
% PROOF
%
\textit{Proof:} 
Follows from Proposition \ref{prop:skron_propts} \ref{prop:skron_propts_mixed_prod}S). 
$\hfill\blacksquare$

These key algebraic properties enable the following fundamental result, the basis of our proposed MEE framework:

% *************************************************************************
%
% THEOREM: SOLUTION OF dEIRL (PRESCALED)
%
% *************************************************************************

\begin{theorem}[MEE Framework \& the dEIRL Algorithm: Modulation Invariance]\label{thm:dEIRL_prescaling_invariance}
$P_{i,j} \in \mathbb{S}^{n_{j}}$, $P_{i,j} > 0$ satisfies the dEIRL regression (\ref{eq:DIRL_nonlin_lsq}) if and only if $\tilde{P}_{i,j} = S_{j}^{-T} P_{i,j} S_{j}^{-1}$ satisfies the MEE regression (\ref{eq:DIRL_nonlin_lsq_prescaled}). Furthermore, the original regression (\ref{eq:DIRL_nonlin_lsq}) and MEE regression (\ref{eq:DIRL_nonlin_lsq_prescaled}) are related by
\begin{align}
	\tilde{\mathbf{A}}_{i,j}
	=
	\mathbf{A}_{i,j} (S_{j} \skron S_{j})^{T},
	\qquad
	\tilde{\mathbf{b}}_{i,j}
	=
	\mathbf{b}_{i,j}.	
	\label{eq:DIRL_nonlin_lsq_Ab_orig_vs_prescaled}
\end{align}

\end{theorem}

% *************************************************************************
%
% PROOF
%
% *************************************************************************
%
\textit{Proof:} 
The first assertion follows immediately from Theorems \ref{thm:Kleinman_prescaling_invariance} and \ref{thm:DIRL_nonlin_Kleinman_equivalence}. The relation (\ref{eq:DIRL_nonlin_lsq_Ab_orig_vs_prescaled}) follows from application of the symmetric Kronecker product algebra developed in Propositions \ref{prop:skron_propts} and \ref{prop:dEIRL_ops_propts}.
%
% ***********************
%
% QED
%
$\hfill\blacksquare$

Theorem \ref{thm:dEIRL_prescaling_invariance} definitively concludes our answer to question 2) posed at the beginning of this section for the dEIRL algorithm and our proposed MEE framework, revealing substantial parallels to the classical Kleinman's algorithm. Crucially, the dEIRL regression matrix $\mathbf{A}_{i,j} \in \mathbb{R}^{l_{j} \times \underline{n}_{j}}$ (\ref{eq:DIRL_nonlin_lsq_A}) is mutliplied on the right by the \emph{same} modulation matrix $(S_{j} \skron S_{j})^{T} \in \GL(\underline{n}_{j})$ to form the MEE regression matrix $\tilde{\mathbf{A}}_{i,j}$ (\ref{eq:DIRL_nonlin_lsq_prescaled}). As is the case with Kleinman's algorithm, the regression target vector $\mathbf{b}_{i,j} \in \mathbb{R}^{l_{j}}$ (\ref{eq:DIRL_nonlin_lsq_b}) remains unchanged under MEE. Furthermore, this vector is given by $\mathbf{b}_{i,j} = - I_{x_{j}, x_{j}} \svec(Q_{i,j})$, which is simply the product of the integral matrix $I_{x_{j}, x_{j}}$ (\ref{eq:I_xy_def}) with the ALE regression vector $- \svec(Q_{i,j})$ (\ref{eq:ALE_svec_Kleinman}).
The parallelisms under which these two algorithms interact with the symmetric Kronecker product algebra developed in this work presents a significant practical advantage to real-world control designers: The same physics-based prescaling insights which readily apply to solving classical control problems may be ported directly to dEIRL's MEE framework. We summarize these key algebraic properties in Table \ref{tb:prescaling_propts}.

%\FloatBarrier

%\fi				% Comment this line and \iffalse to include section

% *************************************************************************
%
% EXPLORATION STUDIES
%
% *************************************************************************

% ***********************
%       
% SUBSECTION: EXPLORATION STUDY I: LINEAR SECOND-ORDER
%     

%\iffalse			% Comment this line and \fi to include section

% ************************************************************************
% ************************************************************************
% ************************************************************************
%
% SECTION: EVALUATION STUDIES
%
% ************************************************************************
% ************************************************************************
% ************************************************************************

\section{Evaluation Studies}\label{sec:ES}

Having developed the algebraic properties of dEIRL's MEE framework, we now demonstrate how MEE may be used as an intuitive, practical tool for real-world designers. We begin with addressing the motivating linear example in first presented in Section \ref{sec:motivating_ex} to illustrate key MEE design principles, and then we apply these insights to a real-world HSV example in Section \ref{ssec:ES_hsv}. In both cases, using little more than physics-based dynamical insights, MEE offers at least an order of magnitude reduction in dEIRL problem conditioning.
These evaluations were performed in MATLAB R2022b, on an NVIDIA RTX 2060, Intel i7 (9th Gen) processor. All numerical integrations in this work are performed in MATLAB's adaptive \texttt{ode45} solver to ensure solution accuracy. Code for the dEIRL algorithm can be found at \cite{BA_Wallace_github_TNNLS_2023:webpage}.

% ************************************************************************
% ************************************************************************
% ************************************************************************
%
% SUBSECTION: EVALUATION STUDIES -- NOMINAL MODEL
%
% ************************************************************************
% ************************************************************************
% ************************************************************************

% ***********************
%
% RELATIVE PATH TO FIGURES
%
\renewcommand{\relpath}{figures/}

\subsection{Evaluation 1: Motivating Example}\label{ssec:ES_lin2d}

\begin{figure*}[ht]
    \centering
    \subfloat[]{\includegraphics[height=\figsize]{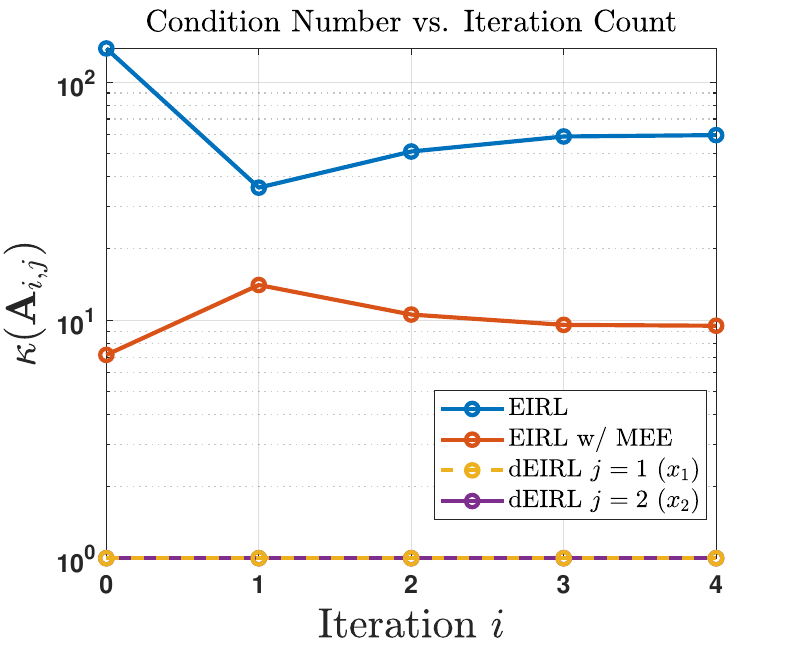}%
    \label{fig:ES_lin2d_cond_A_vs_i}}
    \hfil
    \subfloat[]{\includegraphics[height=\figsize]{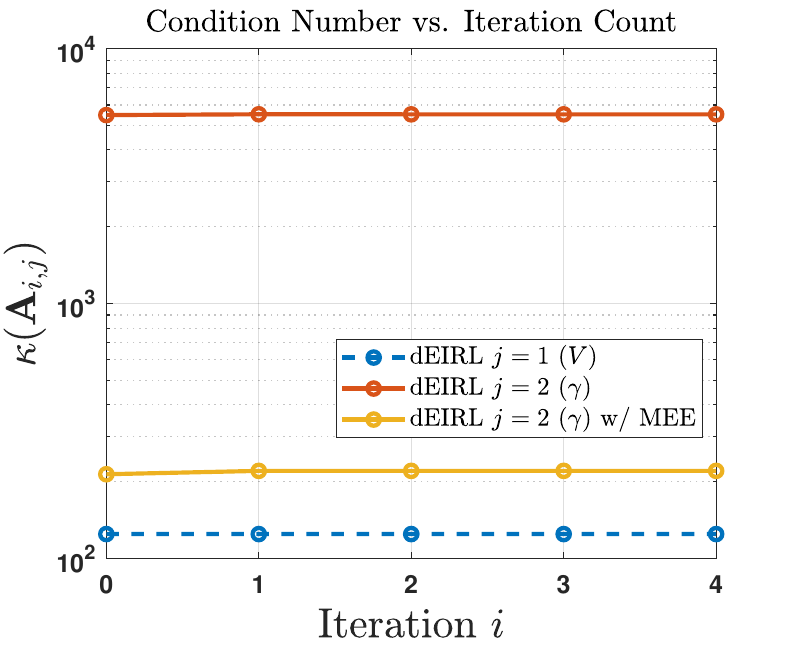}%
    \label{fig:ES_hsv_cond_A_vs_i}}
    \caption{Conditioning number $\kappa(\textbf{A}_{i,j})$ (\ref{eq:DIRL_nonlin_lsq_A}) versus iteration count $i$, with and without prescaling.
    (a): Linear second-order system (Section \ref{ssec:ES_lin2d}).
    (b): HSV system (Section \ref{ssec:ES_hsv}).
    }
    \label{fig:cond_A_vs_i}
\end{figure*}

Consider the motivating linear example (\ref{eq:sys_lin2d}) first discussed in Section \ref{sec:motivating_ex}, with identical hyperparameter selections. We present the resulting peak condition number data in Table \ref{tb:ES_lin2d_cond}, and the corresponding iteration-wise conditioning response in Figure \ref{fig:ES_lin2d_cond_A_vs_i}. 
As noted previously, the EIRL algorithm converges to within $1.62 \times 10^{-9}$ of the optimal $K^{*}$; however, it exhibits large peak condition number of 138.47 (Table \ref{tb:ES_lin2d_cond}). This is caused by saturation constraints in the low-bandwidth loop $j = 2$, which result in a factor of ten separation between the state response $x_{2}(t)$ in the low-bandwidth loop $j = 2$ and the response $x_{1}(t)$ in the high-bandwidth loop $j = 1$.

% *************************************************************************
%
% TABLE: CONDITIONING
%
% *************************************************************************

\begin{table}[h]
	\caption{Eval. 1: Max/Min Conditioning}
	\begin{minipage}{0.5\textwidth}	
	\centering
	\begin{tabular}{|c|c||c|c|}
	        \hline
	        Algorithm & Loop $j$ & $\max\limits_{i}(\kappa(\mathbf{A}_{i,j}))$ & $\min\limits_{i}(\kappa(\mathbf{A}_{i,j}))$ 
	        \\
	        \hline
	        % ***** EIRL *****
	        \hline
	        EIRL & 1 & \cellcolor{black!\lightshade} 138.47 &  36.04
	        \\
	        \hline	      
	        % ***** EIRL WITH PRESCALING *****
	        \hline
	        EIRL & \multirow{2}{*}{1} & \multirow{2}{*}{14.05} & \multirow{2}{*}{7.14} 
	        \\
	        \hhline{|~|~||~|~|}
			w/ MEE &  &  & 
	        \\
	        \hline		  
	        % ***** dEIRL *****
	        \hline
	        \multirow{2}{*}{dEIRL} & 1 & 1.00 & 1.00 
	        \\
	        \hhline{|~|-||-|-|}
			& 2 & 1.00 & 1.00
	        \\
	        \hline	 	        	        
	\end{tabular}
	%\vspace{-0.3cm}	
	\end{minipage}	
	\label{tb:ES_lin2d_cond}
\end{table}

Intuition offers a clear solution: The state response $x_{2}(t)$ in the low-bandwidth loop needs to be scaled up by a factor of ten to improve scaling. This is precisely where MEE offers immense practical numerical benefits to designers. Indeed, choosing the natural modulation matrix $S = \texttt{diag}(1, 10) \in \GL^{+}(2)$ drastically improves EIRL conditioning, reducing it by a factor of ten from 138.47 before MEE to 14.05 after MEE (Table \ref{tb:ES_lin2d_cond}), a reduction seen iteration-wise across the board (Figure \ref{fig:ES_lin2d_cond_A_vs_i}). Thus, using little beyond common-sense principles, MEE can offer conditioning reductions of an order of magnitude to designers using the EIRL/dEIRL algorithm, cementing this framework's already substantial numerical performance guarantees \cite{\TNNLSdEIRLCitation}.

We conclude this section by employing a decentralized design (i.e., dEIRL) in each of the $N = 2$ loops. Using identical hyperparameters, the resulting final controllers $K_{i^{*}, 1}$, $K_{i^{*}, 2}$ converge to within $1.38 \times 10^{-11}$ and $1.49 \times 10^{-9}$ of the optimal controllers $K_{1}^{*}$, $K_{2}^{*}$ in each loop, respectively. Furthermore, dEIRL has unity conditioning in each loop (since the dimension of each is $\underline{n}_{1} = \underline{n}_{2} = 1$), illustrating the general principle that dEIRL's use of physically-motivated dynamical insights enables further learning performance improvements.

% *************************************************************************
%
% FIGURE: WEIGHT RESPONSES
%
% *************************************************************************

\renewcommand{\relpathone}{lin2d/}	
\renewcommand{\relpathtwo}{hsv/}

%\FloatBarrier

%\fi				% Comment this line and \iffalse to include section

% ***********************
%       
% SUBSECTION: CASE STUDY II: HSV
%     

%\iffalse			% Comment this line and \fi to include section

% ************************************************************************
% ************************************************************************
% ************************************************************************
%
% SUBSECTION: EVALUATION STUDIES -- 25% DRAG COEFFICIENT MODELING ERROR
%
% ************************************************************************
% ************************************************************************
% ************************************************************************

% ***********************
%
% RELATIVE PATH TO FIGURES
%
\renewcommand{\relpath}{figures/ES_CL_error/}

\subsection{Evaluation 2: Hypersonic Vehicle (HSV) Example}\label{ssec:ES_hsv}

Having now motivated the significant numerical benefits of MEE on an illustrative example in Section \ref{ssec:ES_lin2d}, we now demonstrate how these principles may be readily applied to a real-world nonlinear, nonminimum phase HSV system. The HSV model considered was developed from NASA Langley aeropropulsive data \cite{Shaughnessy_Pinckney_McMinn_Cruz_Kelley_HSV_modeling_NASA_Langley:1990} and has proven a standard testbed for seminal works such as \cite{Marrison_Stengel_HSV_tracking_robust:1998,Wang_Stengel_HSV_tracking_robust:2000,Xu_Mirmirani_Ioannou_HSV_tracking_neural_adaptive:2003,Xu_Mirmirani_Ioannou_HSV_tracking_adaptive_sliding_mode:2004}. 
We offer a complete analysis of the model in the original dEIRL work \cite{\TNNLSdEIRLCitation}, so we omit the dynamical equations and discussion here for sake of brevity. In sum, the HSV is fifth-order, with states $x = \left[ V, \, \gamma, \, \theta, \, q, \, h \right]^{T}$, where $V$ is the vehicle airspeed, $\gamma$ is the flightpath angle (FPA), $\theta$ is the pitch attitude, $q$ is the pitch rate, and $h$ is the altitude. The controls are $u = \left[ \delta_{T}, \, \delta_{E} \right]^{T}$, where $\delta_{T}$ is the throttle setting, and $\delta_{E}$ is the elevator deflection. We examine the outputs $y = \left[ V, \, \gamma \right]^{T}$. The HSV is naturally a two-loop system consisting of the weakly-coupled velocity subsystem $j = 1$ (associated with the airspeed $V$ and throttle control $\delta_{T}$) and rotational subsystem $j = 2$ (associated with the FPA $\gamma$, attitude $\theta, q$, and elevator control $\delta_{E}$). For decentralized design, we augment the plant at the output with the integrator bank $z = \int y \, d\tau = \left[ z_{V}, \, z_{\gamma}  \right]^{T} = \left[ \int V \, d\tau, \, \int \gamma \, d\tau \right]^{T}$. The state/control vectors are thus partitioned as $x_{1} = \left[ z_{V}, \, V \right]^{T}$, $u_{1} = \delta_{T}$ ($n_{1} = 2$, $m_{1} = 1$) and $x_{2} = \left[ z_{\gamma}, \, \gamma, \, \theta, \, q \right]^{T}$, $u_{2} = \delta_{E}$ ($n_{2} = 4$, $m_{2} = 1$).

Running dEIRL with identical hyperparameter selections to those enumerated in \cite[Section \TNNLSdEIRLSecESSetup]{\TNNLSdEIRLCitation}, the resulting final controllers $K_{i^{*}, 1}$, $K_{i^{*}, 2}$ converge to within $1.07 \times 10^{-6}$ and $2.85 \times 10^{-5}$ of the optimal controllers $K_{1}^{*}$, $K_{2}^{*}$ in each loop, respectively -- a significant synthesis guarantee for this real-world aerospace example. We include the max/min conditioning data in Table \ref{tb:ES_hsv_cond} and corresponding conditioning response in Figure \ref{fig:ES_hsv_cond_A_vs_i}. As a technical note, the numerical conditioning data presented here varies slightly from that of the original dEIRL study \cite{\TNNLSdEIRLCitation} due to our re-scaling the map $\svec$ (\ref{eq:svec_def}) to make this operator an isometry (cf. Proposition \ref{prop:skron_HS_iso}).

% *************************************************************************
%
% TABLE: CONDITIONING
%
% *************************************************************************

\begin{table}[h]
	\caption{Eval. 2: Max/Min Conditioning}
	\begin{minipage}{0.5\textwidth}	
	\centering
	\begin{tabular}{|c|c||c|c|}
	        \hline
	        Algorithm & Loop $j$ & $\max\limits_{i}(\kappa(\mathbf{A}_{i,j}))$ & $\min\limits_{i}(\kappa(\mathbf{A}_{i,j}))$ 
	        \\
	        \hline
	        % ***** dEIRL *****
	        \hline
	        \multirow{2}{*}{dEIRL} & 1 & 124.38 & 124.36
	        \\
	        \hhline{|~|-||-|-|}
			& 2 & \cellcolor{black!\lightshade} 5517.97 & \cellcolor{black!\lightshade} 5478.64
	        \\
	        \hline	       
	        % ***** dEIRL WITH PRESCALING *****
	        \hline
	        dEIRL & \multirow{2}{*}{2} & \multirow{2}{*}{220.13} & \multirow{2}{*}{213.53} 
	        \\
	        \hhline{|~|~||~|~|}
			w/ MEE &  &  & 
	        \\
	        \hline		  	        	        
	\end{tabular}
	%\vspace{-0.3cm}	
	\end{minipage}	
	\label{tb:ES_hsv_cond}
\end{table}

Examination of Table \ref{tb:ES_hsv_cond} shows that worst-case conditioning is already acceptable in the velocity loop $j = 1$ at 124.38. Thus, no modulation $S_{1} = I_{2}$ in loop $j = 1$ is necessary. However, conditioning in the higher-dimensional, unstable, nonminimum phase FPA loop $j = 2$ is worse at 5517.97. Although this represents a substantial reduction of fourteen orders of magnitude from prevailing ADP-based CT-RL methods \cite{\TNNLSdEIRLCitation,BA_Wallace_J_Si_CT_RL_review:2022}, conditioning reductions in this loop are still desired for real-world numerical reliability. Furthermore, just as in the motivating example studied in Section \ref{ssec:ES_lin2d}, a few minutes of investigation yields a physically-intuitive explanation of the cause of the conditioning issue. Within the FPA loop $j = 2$ is the FPA subsystem $\gamma$ itself (stable, nonminimum phase), alongside the attitude subsystem $\theta, q$ (unstable, minimum phase). The FPA dynamics have a bandwidth roughly a decade below that of the attitude dynamics. As a result, the pitch $\theta$ generally exhibits larger responses than the FPA, and the pitch rate $q$ by virtue of differentiation magnifies this response amplitude discrepancy. 

As in the simple linear example, the designer course of action is clear here: The attitude states $\theta, q$ need to be scaled down to equilibrate their amplitudes with that of the FPA response $\gamma$ and thereby improve scaling in the regression matrix $\mathbf{A}_{i,2}$ (\ref{eq:DIRL_nonlin_lsq_A}). Generally, it is common for angular state variables to be expressed in degrees for the sake of flight control implementation \cite{Stengel_flight_dynamics:book:2022,Dickeson_AA_Rodriguez_Sridharan_etal_HSV_decentralized_control:2009,AA_Rodriguez_Dickeson_etal_HSV:2008,AA_Rodriguez_Pradhan_BA_Wallace_etal_hawkmoth_MAV:2020}. Thus, a remedy a designer may likely choose is to simply convert the pitch $\theta$ and pitch rate $q$ to radians for the purposes of the MEE regression (\ref{eq:DIRL_nonlin_lsq_prescaled}), while keeping the FPA $\gamma$ and integral augmentation $z_{\gamma}$ in degrees: $S_{2} = \texttt{diag}(1, 1, \pi/180, \pi/180) \in \GL^{+}(4)$. After the MEE regression is complete, the pitch $\theta$ and pitch rate $q$ may then be converted back to degrees for control implementation via the inverse transformation $S_{2}^{-1}$in (\ref{eq:Pij_Kij_vs_tPij_tKij}) while preserving the convergence/stability of the resulting controller, a result guaranteed by the MEE framework in Theorem \ref{thm:dEIRL_prescaling_invariance}. 
We include this MEE conditioning data in the FPA loop $j = 2$ in Table \ref{tb:ES_hsv_cond} and Figure \ref{fig:ES_hsv_cond_A_vs_i}. As can be seen, this simple radians/degrees conversion reduces worst-case conditioning by factor of 25 from 5517.97 without MEE to 220.13 with MEE, a conditioning reduction observed iteration-wise across the board in Figure \ref{fig:ES_hsv_cond_A_vs_i}. In light of the higher dimension and dynamical challenges associated with the FPA loop $j = 2$, a near equalization of the conditioning in this loop with that of the velocity loop $j = 1$ is a substantial real-world numerical result.

Whereas in our previous study we illustrated the motivation, method, and results of MEE on a simple academic example, here we show definitively that the same first-principles intuitions of the dynamics may be extended to MEE on significant, challenging practical applications -- with potentially even greater factors of performance improvement. 
We demonstrate how MEE may be used systematically in conjunction with decentralization and multi-injection, equipping designers with an unrivaled suite of practical numerical capabilities. 

%The exhibited numerical performance improvements are even greater for the real-world HSV than for the academic example, proving prescaling an intuitive and powerful tool for real-world designers to use on the already substantial dEIRL framework. 

%\FloatBarrier

%\fi				% Comment this line and \iffalse to include section

% *************************************************************************
%
% CONCLUSION
%
% *************************************************************************

%\iffalse			% Comment this line and \fi to include section

% ************************************************************************
% ************************************************************************
% ************************************************************************
%
% CONCLUSION
%
% ************************************************************************
% ************************************************************************
% ************************************************************************

\section{Conclusion \& Discussion}\label{sec:conclusion}

This work presents a novel modulation-enhanced excitation (MEE) framework to address fundamental PE issues in continuous-time reinforcement learning control. We apply this MEE framework to the cutting-edge suite of EIRL algorithms, enabling numerical performance enhancements while preserving their key convergence/stability guarantees via new symmetric Kronecker product algebra. 
%MEE is demonstrated to significantly improve EIRL excitation, resulting in substantial conditioning reductions in the learning regression. 
Using simple design principles, MEE is demonstrated to improve conditioning properties of dEIRL by at least an order of magnitude in numerical studies -- by a factor of 25 on the significant real-world hypersonic vehicle example. When MEE is combined with the multi-injection and decentralization of dEIRL, this method now offers a three-pronged designer approach for maximizing algorithm numerical performance, enabling control synthesis results unprecedented in CT-RL \cite{BA_Wallace_J_Si_CT_RL_review:2022,\TNNLSdEIRLCitation}.

To enable the MEE framework, we present novel results on the symmetric Kronecker product \cite{Alizadeh_Haeberly_Overton_skron:1998,Todd_Toh_Tutuncu_skron:1998,Tuncel_Wolkowicz_skron:2005,Kalantarova_Tuncel_skron:2021,Kalantarova_PhD_thesis_skron:2019}. 
%We develop symmetric Kronecker product algebra on rectangular matrices, and we show new algebraic/spectral properties in the previously-addressed square-matrix case \cite{Alizadeh_Haeberly_Overton_skron:1998,Todd_Toh_Tutuncu_skron:1998,Tuncel_Wolkowicz_skron:2005,Kalantarova_Tuncel_skron:2021,Kalantarova_PhD_thesis_skron:2019}. 
This work also motivates the concept of the symmetric Kronecker sum, which we demonstrate is the natural analogue to its standard counterpart in its algebraic, spectral, and exponentiation properties, as well as its central role in solving ALEs.

\bibliographystyle{IEEEtran}
\bibliography{refs}
%\bibliography{\bibpath refs}
%\bibliography{"C:/Users/Brent/AppData/Local/Programs/MiKTeX 2.9/bibtex/bib/00 Brent Bib/2022-12-14/refs.bib"}

%\printbibliography

%\fi					% Uncomment this and \iffalse to skip section

\vfill

\end{document}